\documentclass[pra,aps,twocolumn,10pt,notitlepage,longbibliography]{revtex4-2}
\usepackage{graphicx}
\usepackage{physics}
\usepackage[capitalise]{cleveref}
\usepackage{tikz}
\usetikzlibrary{patterns,arrows,calc,decorations.pathmorphing, shapes.geometric,backgrounds}

\begin{document}
\title{Analytic solution to the nonlinear generation of squeezed states in a thermal bath}
\author{Paul R. B. Hughes}
\email{p.hughes@queensu.ca}
\author{Marc M. Dignam}
\affiliation{Department of Physics, Engineering Physics and Astronomy,
Queen's University, Kingston, ON K7L 3N6, Canada}

\begin{abstract}
We model squeezed state generation in a lossy optical cavity in the presence of a thermal bath using the Lindblad master equation. We show that the exact solution is a squeezed thermal state, where thermal photons arise both from loss and from the thermal bath. We derive an exact, closed-form solution for the evolution of the quadrature uncertainty arising from pulsed degenerate spontaneous parametric down conversion in the cavity. We apply this solution under different pump conditions and show in detail how the thermal environment reduces quadrature squeezing as well as the second order coherence function. 
\end{abstract}

\maketitle

\textit{Introduction.}  
Nonlinear optical processes such as spontaneous parametric down conversion (SPDC) and spontaneous four wave mixing are often used to generate nonclassical states of light, such as photon pairs, single-mode quadrature squeezed states, multimode squeezed states, and entangled optical modes \cite{milburnOPO,WMQO,WallsD.F1983Ssol,PhysRevLett.57.2520,RevModPhys.82.1155}. Photon pairs can be used as a heralded single-photon source or as entangled two-photon states \cite{PhysRevA.65.032323,PhysRevA.109.053711}.  Quadrature squeezed states can be used to reduce the uncertainty in interferometric measurements \cite{GrangierP.1987Sepi,ligo}
or to create continuous variable entanglement \cite{CVonchip,PhysRevA.109.040101}. Because the nonlinear interactions are generally quite weak, one usually requires a resonator to form an optical parametric oscillator (OPO) to enhance the process. Some common resonators are microring resonators \cite{colinMRR,Yang:07,Dutt:16} and Fabry-Perot cavities \cite{PhysRevLett.57.2520,Yurke1984}. The use of a resonator has been shown to lower the uncertainty in one quadrature below vacuum fluctuations at the expense of the other, but with a steady-state squeezing limit of 3dB in the resonator \cite{milburnOPO,hossein}. Larger squeezing can been obtained for light coupled out of the resonator \cite{Yurke1984,Schneider1987,Wu1987,Aoki2006,Takeno2007,Vahlbruch2008,Mehmet2011}, but within it, alternative methods are required to overcome the 3dB limit. These include pulsed excitation \cite{colinMRR}, quantum feedback \cite{RevModPhys.52.341,PhysRevB.71.235407,PhysRevLett.117.100801}, and dissipation \cite{KronwaldAndreas2013Alsb,PRXQuantum.2.020323}.

The steady-state quantum fluctuations of the state in an OPO can be derived using the Langevin equations for a stochastic process \cite{doi:10.1080/713820531,WMQO,PhysRevA.45.1804,PhysRevA.30.1386}. When loss and detuning are not considered in these systems, the signal field in the OPO is a squeezed vacuum state. Other groups have considered the effects of detuning from resonance on squeezing in the OPO in the steady state \cite{DunlopA.E.2006Goaf,Jabri:19}. The effects of loss and a thermal bath on the generation and evolution of the density operator of the light in an OPO can be modeled using the Lindblad master equation (LME). Recently it was shown that when there is no thermal bath, the exact solution to the LME is a squeezed thermal state (STS) \cite{hossein}. 

At optical or near infrared frequencies, the thermal effects of an environment on the generation and nature of the squeezed states is negligible at or below room temperature when employing SPDC in a resonator. At lower frequencies of a few tens of terahertz or less, thermal noise can have a significant effect on the generation, evolution, and final state. In particular quadrature squeezing can be greatly reduced unless one cools the system to millikelvin temperatures \cite{PRXQuantum.2.020323}.  It is therefore important to be able to accurately and efficiently model the evolution dynamics in such systems and to quantify the effects of temperature on the final squeezed state for CW and pulsed pumping configurations. To this end, in this work, we derive the exact solution to the LME for SPDC in a lossy OPO coupled to a thermal bath and show that the density operator is that of a STS. With this exact solution, we are able to derive a closed-form solution for the evolution of the quadrature uncertainty for an arbitrary, un-chirped pump pulse and to examine the evolution of the squeezing parameter, the thermal photon number and the second-order quantum coherence function.

The paper is organized as follows.  
We first outline the theory behind the generation of the signal field in a resonator. We show that the solution to the LME is a STS, where the thermal photon number and squeezing parameter evolution are described by three coupled first-order differential equations. Using these equations, we derive closed form solutions for the quadrature uncertainties, which to the best of the authors' knowledge have never been derived previously. 
Next, we examine the transient and steady-state properties of the system excited by a constant-amplitude pump pulse, presenting an exact analytic solution for the uncertainties. Using the second order quantum coherence and the quadrature uncertainty, we investigate the nonclassicallity of the light and discuss the threshold where the quadrature is squeezed below vacuum noise. Finally, 
we present the results for a Gaussian pump pulse and examine the relationship between the pulse amplitude and the quadrature squeezing as a function of the bath temperature.

\begin{figure}[h]
    \resizebox{0.95\columnwidth}{!}{
    \hskip-0.8ex  
    \begin{tikzpicture}[background rectangle/.style={fill=white}, show background rectangle]
\draw [line width = 2, decorate, decoration={snake}, -](0, 0) -- node [above left] {\Large $\hbar\omega_p$} (3, 0) -- node [above] {\Large $\alpha(t)$} node [below] {\Large $\gamma$} (3.6, 0) -- (4.5, 0);

\path (2, -2) node {\Large $T_b$};

\def\rad{2.6}
\draw [line width = 2] (4, 0) arc (180:150:\rad);
\draw [line width = 2] (4, 0) arc (180:210:\rad);
\draw [line width = 2] (8.5, 0) arc (0:30:\rad);
\draw [line width = 2] (8.5, 0) arc (0:-30:\rad);

\draw [line width = 1, decorate, decoration={snake}, -](4.3, 0.5)--
  node [above] {\Large $\hbar\omega$}(8.2, 0.5);
\draw [line width = 1, decorate, decoration={snake}, -](8.2, -0.5)--
  node [below] {\Large $\hbar\omega$}(4.3, -0.5);

\draw [line width = 1, decorate, decoration={snake}, -](7.4, -1)--
  node [below right] {\Large $\Gamma$}(2.5, -2);

\path (4.55, 0) node [dart, fill=black, scale=0.75] {};
\path (8.22, 0.5) node [dart, fill=black, scale=0.5] {};
\path (4.28, -0.5) node [dart, fill=black, scale=0.5, shape border rotate=180] {};
\path (7.38, -1.02) node [dart, fill=black, scale=0.5, rotate=25] {};
\path (2.45, -2) node [dart, fill=black, scale=0.5, rotate=-155] {};
\end{tikzpicture}
    \hskip-1.3ex}
    \caption{Schematic diagram of the system. A coherent state $\alpha(t)$ at frequency $\omega_p$ pumps the resonator which is coupled $(\Gamma)$ to the environment at temperature $T_b$, generating pairs of signal photons at $\omega$.}
    \label{schematic}
\end{figure}
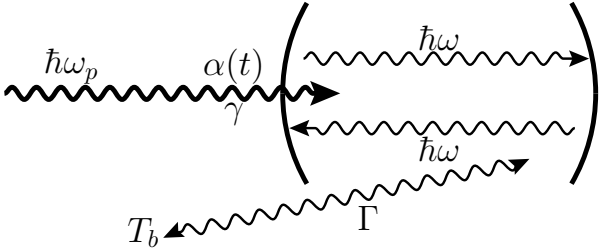

\textit{Theory.}
We consider the generation of a squeezed state in a single mode of a resonant cavity with frequency $\omega$. Shown schematically in \cref{schematic}, the system consists of a resonant cavity that is coupled to a thermal bath of photons and is excited by a coherent optical pump.
The pump operates at a frequency $\omega_p = 2\omega$ and generates signal photons in the resonator through SPDC. 
The pump is a coherent state with time-dependent coherent state amplitude $\alpha(t) = \alpha_0(t) e^{-i\omega_p t}$, where $\alpha_0(t)$ is the pump envelope. Thus, we treat the pump classically in the undepleted pump approximation. When the interaction with the bath is neglected, the system Hamiltonian is given by \cite{hossein}
\begin{equation}
H = \hbar\omega {b}^\dag b + \alpha(t)\gamma {b^\dag}^2 + \alpha^*(t) \gamma^* b^2,
\label{hamm}
\end{equation}
where $b^\dag$ ($b$) is the creation (annihilation) operator of photons in the cavity and $\gamma = \hbar \omega_p \chi_{eff}^{(2)}/n_{eff}^2$ is the coupling coefficient of the pump field to the signal field for an effective second order nonlinear susceptibility $\chi_{eff}^{(2)}$ and refractive index $n_{eff}$ in the cavity. 

The signal mode in the cavity is coupled to a thermal bath at temperature $T_b$, which has a mean photon number, $n_b = (\exp(\hbar\omega/kT_b) - 1)^{-1}$ at the signal frequency. The density operator of the cavity $\rho(t)$ evolves according to the LME \cite{OpenQuantum}
\begin{align}
\begin{aligned}
\frac{d}{dt}\rho (t) = -\frac{i}{\hbar}\comm{H}{\rho(t)} &+ \Gamma (n_b+1) D[{b}](\rho(t)) \\ &+ \Gamma n_b D[{b}^\dag](\rho(t)),
\label{master}
\end{aligned}
\end{align}
where $\Gamma$ is the power decay constant of the cavity photons into the bath, while 
\begin{equation}
D[F](\rho) \equiv F\rho F^\dag - \frac{1}{2}\acomm{F^\dag F}{\rho}    
\end{equation} 
is the dissipator, which accounts for the two-way coupling between bath and cavity.

In a previous work, it was shown that for the special case where the bath is at zero temperature ($n_b=0$), the exact solution to \cref{master} is a STS \cite{hossein}. In this work, we prove that at non-zero temperatures, the solution is still a STS, but that the evolution of the squeezing and thermal temperature depends, in general, on the bath temperature.

We find that as long as the initial state is the vacuum, a thermal state, or a STS, the exact solution to the above LME is the time-dependent STS:
\begin{gather}
\rho(t) = S(\xi(t))\rho_T(n_{th}(t))S^\dag(\xi(t)),
\label{STS}
\end{gather}
where
\begin{equation}
S(\xi) = \exp\left[ \frac{1}{2}(\xi^* b^2 - \xi {b^\dag}^2)\right] \label{squeezeOP}
\end{equation} 
is the squeezing operator, with the time-dependent, complex squeezing factor $\xi(t) = u(t)e^{i\phi(t)}$, and  
\begin{gather}
\rho_T(n_{th}) = \frac{1}{1 + n_{th}} \left( \frac{n_{th}}{1 + n_{th}} \right)^{b^\dag b}
\end{gather}
is a thermal state, with a time-dependent thermal population, $n_{th}\left(t\right)$.

To show that this is the exact solution, we write the density operator in the form
\begin{gather}  
\rho(t) = S(\xi) \rho_T^{1/2}(n_{th}) O(t) \rho_T^{1/2}(n_{th}) S^\dag(\xi).
\end{gather}
We then need to prove that the operator, 
\begin{gather}
O(t) = \rho_T^{-1/2}(n_{th})S^\dag(\xi)\rho(t)S(\xi)\rho_T^{-1/2}(n_{th}) \label{Oeqn}
\end{gather}
is simply the identity operator for all time. In the supplementary material, we show that this is indeed the case as long as the thermal photon number, the squeezing amplitude, and phase evolve according to the following three coupled first-order differential equations:
\begin{gather}
\frac{dn_{th}}{dt} = \Gamma\left[n_b \cosh(2u) + \sinh^2(u) - n_{th} \right], \label{nthfirst} \\
\frac{du}{dt} = -\frac{i}{\hbar}(\gamma^* \alpha^* e^{i\phi} - \gamma \alpha e^{-i\phi}) - \frac{\Gamma}{2}\sinh(2u) \frac{2n_b + 1}{2n_{th} + 1}, \label{u0} \\
\frac{d\phi}{dt} = -2\omega + \frac{2}{\hbar}\frac{\cosh(2u)}{\sinh(2u)}(\gamma^* \alpha^* e^{i\phi} + \gamma \alpha e^{-i\phi}). \label{phi0}
\end{gather}

\cref{nthfirst,u0,phi0}, are the dynamic equations valid for any initial STS and for any $\alpha(t)$. In all that follows, we restrict ourselves to unchirped pump pulses, such that $\alpha_0(t)\gamma = |\alpha_0(t)\gamma|e^{i\theta}$, where $\theta$ is a time-independent phase and we assume that the initial state is an unsqueezed thermal state, such that $u(0) = 0$. To avoid a divergence on the left hand side of \cref{phi0} that arises at $t=0$ for an unsqueezed state, we impose the initial condition on $\phi$ that $(\gamma^* \alpha^*(0) e^{i\phi(0)} + \gamma \alpha (0)e^{-i\phi(0)}) = 0$, or $\phi(0) = \theta + \pi/2$. Using this in \cref{phi0}, we see that for all time $(\gamma^* \alpha^*(t) e^{i\phi(t)} + \gamma \alpha (t)e^{-i\phi(t)}) = 0$ and $\phi (t)=\theta +\pi/2-2\omega t$. \\

We now define the pump function, 
\begin{equation}
g(t) \equiv 4|\alpha_0(t) \gamma|/\hbar \Gamma,    
\end{equation}
which is the ratio of the pumping rate to the loss rate, such that $g(t)=1$ is the critical pump rate at which the injection of photons is exactly balanced by the loss.  Using this definition and the above initial condition, the equations of motion become
\begin{gather}
\frac{dn_{th}}{dt} = \Gamma\left[n_b \cosh(2u) + \sinh^2(u) - n_{th} \right], \label{nth} \\
\frac{du(t)}{dt} = \frac{\Gamma g(t)}{2} - \frac{\Gamma\sinh(2u)}{2}\frac{2n_b + 1}{2n_{th} + 1}, \label{uDE} 
\end{gather}
with
\begin{equation}
\phi (t)=\theta + \pi/2-2\omega t .    
\end{equation}

The above dynamic equations contain an explicit dependence on the temperature of the bath; however, in the limit that $n_b = 0$, they reduce to what we obtained in our previous $T_b=0$ work \cite{hossein}. Additionally, in the simple case that there is no pump present, but the initial state is a thermal state that is not at the bath temperature, these equations show that the system remains a thermal state with the thermal photon number evolving as $n_{th}(t) = n_b + (n_{th}(0) - n_b)e^{-\Gamma t}$, a result that has been shown previously using the LME \cite{StatQuantumOptics}.

We can see from \cref{uDE} that the bath population increases the decay rate of the squeezing factor $u(t)$. However, this contribution is not present for the typical initial state in which the system is in equilibrium with the environment. To see this, let $n_{th}^0$ be the thermal population that would arise if $T_b=0$. We define it using the equation 
\begin{gather}
2n_{th} + 1 = (2n_b + 1)(2n_{th}^0 + 1). \label{nthform}
\end{gather}
Using this in \cref{nth}, we find that
\begin{gather}
\frac{dn_{th}^0}{dt}  = \Gamma[\sinh^2(u) - n_{th}^0], \label{nth0}
\end{gather}
which is exactly the evolution of the thermal photon number when the bath temperature is zero. We can also rewrite \cref{uDE} using $n_{th}^0$ as
\begin{gather}
\frac{du(t)}{dt} = \frac{\Gamma g(t)}{2} - \frac{\Gamma \sinh(2u)}{2(2n_{th}^0 + 1)}. \label{u2}
\end{gather}
The bath population is still implicitly present in these equations, since from \cref{nthform}, the initial value of $n_{th}^0$ is given by
\begin{gather}
n_{th}^0(0) = \frac{n_{th}(0) - n_b}{2n_b + 1}. \label{nth0ic}
\end{gather}
However, in the case where the cavity begins in a thermal state in equilibrium with the bath ($n_{th}(0) = n_b$), the evolution of $n_{th}^0$ and thus squeezing factor $u$ are independent of the bath temperature, while the actual thermal population only depends on $n_b$ through the prefactor of $2n_b + 1$ (see \cref{nthform}) \footnote{We can see from \cref{nth0ic} that these properties arise not just for an initial state in equilibrium with the thermal bath, but any initial thermal population that satisfies $n_{th}(0) = n_b + a(2n_b + 1)$. The dynamics of the squeezing amplitude will then be independent of the bath, but the decay will scale by the arbitrary factor $a$.}. We will examine the dependence of the thermal population, the total population, and the squeezing amplitude on the bath temperature and initial thermal population in more detail 
later in this letter.

We now define the two quadrature operators, 
\begin{gather}
X = b^\dag e^{-i\beta(t)} + b e^{i\beta(t)}, \\
Y = -i(b^\dag e^{-i\beta(t)} - b e^{i\beta(t)}),
\end{gather}
where $\beta (t)$ is the local oscillator phase.  For $\beta(t) \equiv \omega t$, the system is squeezed in $X$ and antisqueezed in $Y$. For a STS, the uncertainties in these quadratures can be shown to be given by $\Delta X^2 = (2n_{th} + 1)e^{-2u}$ and $\Delta Y^2 = (2n_{th} + 1)e^{2u}$ \cite{properties}, which allows us to determine the evolution of the squeezing from the evolution of $n_{th}$ and $u$. Alternatively, we can derive differential equations for the quadrature uncertainties. Taking the time derivative of $\Delta X^2$ and using \cref{nth,uDE} and the hyperbolic identities, we find that
\begin{align}
&\frac{d}{dt} \Delta X^2 = \Big[2\frac{dn_{th}}{dt} - 2(2n_{th} + 1) \frac{du}{dt} \Big]e^{-2u} \nonumber
\\
&\begin{aligned}
= \Gamma\Big[&(2n_b + 1)\left( \cosh(2u) + \sinh(2u)\right)e^{-2u} \\ &- \left(1 + g(t)\right)(2n_{th} + 1)e^{-2u}\Big],
\end{aligned}
\end{align}
which can also be written as
\begin{gather}
\frac{d}{dt}\Delta X^2 = \Gamma \left[ (2n_b + 1) - (1 + g(t))\Delta X^2 \right]. \label{xDE}
\end{gather}
Similarly, 
\begin{gather}
\frac{d}{dt}\Delta Y^2 = \Gamma \left[ (2n_b + 1) - (1 - g(t))\Delta Y^2 \right].\label{yDE}
\end{gather}
These equations show that the squeezing dynamics depend only on the pumping strength and the thermal bath population. Furthermore, the evaluation of the quadrature squeezing only requires the solution of a single first-order differential equation, which is directly solvable using standard techniques. To the authors' knowledge, this is the first time a closed-form solution has been derived for the nonlinear generation and evolution of the quadrature uncertainty in a lossy cavity. 

\textit{Constant Pump and Steady State.}
In this section, we examine the early-time evolution  and steady-state solution of the system when it is excited by a pump that has a constant strength, $g_0$ that is turned on at $t=0$. \\
Before analysing the quadrature uncertainties directly, we return to the question of the dependence of the STS parameters on the bath temperature and initial state of the system. Recall that for the special case where $n_{th}(0)=n_b$, the squeezing factor $u(t)$ is independent of $n_b$. As we now show, when the initial state is a thermal state at a different temperature from the environment, $u(t)$ is still only weakly dependent on both the initial thermal population and the bath temperature.

\begin{figure}[htb]
\includegraphics[width=\columnwidth]{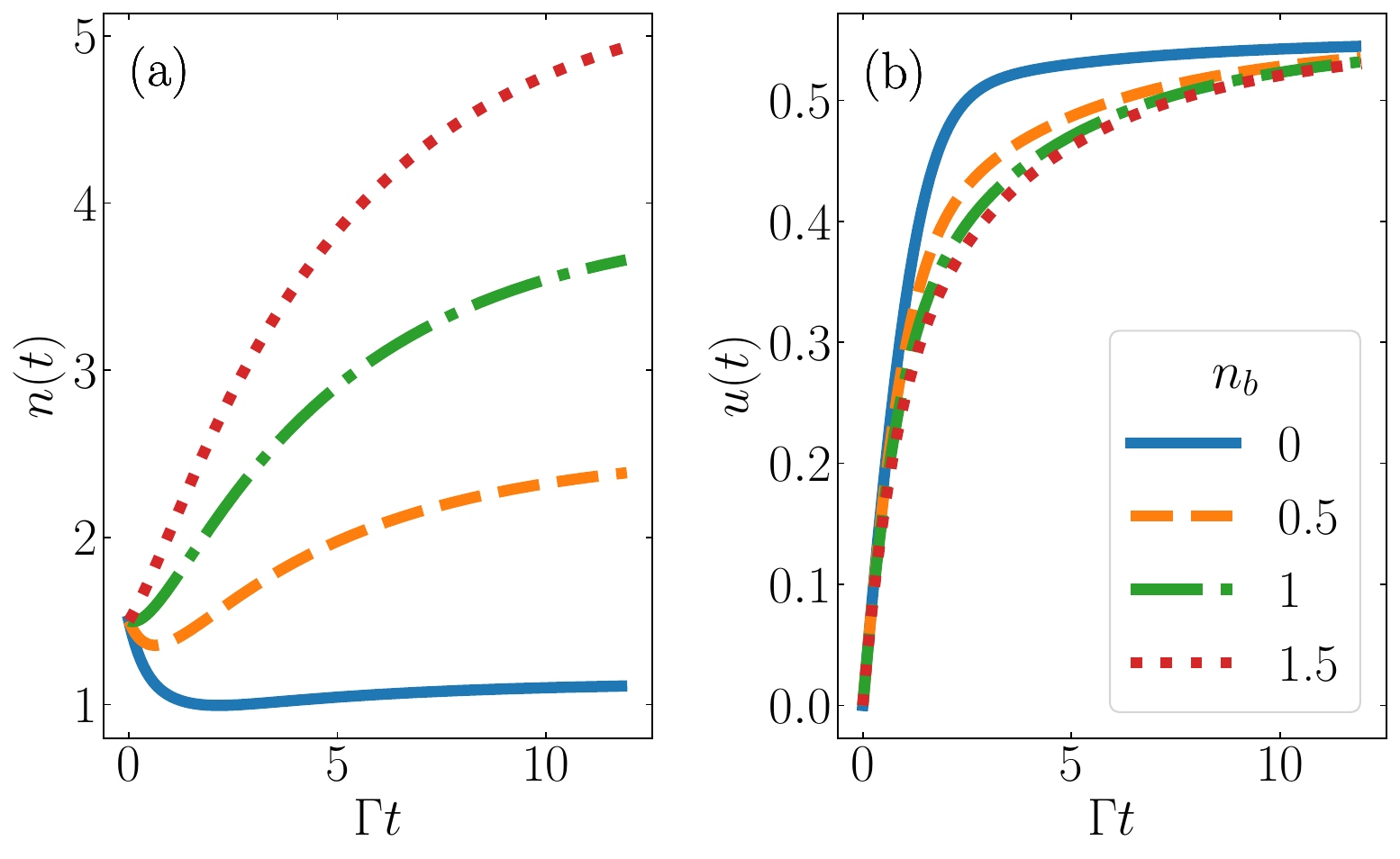}
\caption{(a) Total population $n(t)$ and (b) squeezing factor $u(t)$ as a function of time for an initial thermal state with $n_{th}(0) = 1.5$. The cavity is pumped with a continuous wave pulse of strength $g(t) = g_0 = 0.8$. The different curves are for different thermal bath populations, $n_b$. }\label{cwplot}
\end{figure}

In \cref{cwplot}, we plot the total photon number and squeezing amplitude as a function of time for a cavity pumped by a continuous pulse excitation. It is prepared in the same initial state each time but coupled to environments at different bath temperatures. From \cref{cwplot}(a), we see that when $n_b$ is increased, the photon number increases by much more than simply $n_b$. Meanwhile, \cref{cwplot}(b) shows that $u(t)$ exhibits a small dependence on $n_b$ at early times and that even this dependence disappears as $t\rightarrow \infty$.
We can determine the steady-state characteristics of the continuous wave pump by setting the derivatives in \cref{uDE,nth} to zero. For the steady-state squeezing amplitude, we obtain
\begin{gather}
u^{ss} = \frac{1}{2} \tanh^{-1}(g_0), \label{uss}
\end{gather}
which is independent of the environment and only exists for pumping below the critical pump strength, $g_0 = 1$. The steady-state thermal and total populations are given by
\begin{align}
\begin{aligned}
n_{th}^{ss} &= n_b + \sinh^2(u)(2n_b + 1) \\ &= \frac{1}{2}\left( \frac{2n_b + 1}{\sqrt{1 - g_0^2}} - 1\right), \label{nthss}
\end{aligned}
\end{align}
and
\begin{gather}
n^{ss} = \frac{2n_b + g_0^2}{2(1 - g_0^2)},
\end{gather}
where, to determine the total population, we have used the relation for a STS that $n = n_{th}\cosh(2u) + \sinh^2(u)$ \cite{properties}. 
Thus, as discussed earlier, the thermal environment adds many more photons to the system than just $n_b$, but it does nothing to the steady state squeezing factor.

We now consider the evolution of quadrature uncertainties. The dynamic equations, \cref{xDE,yDE}, can be solved exactly for an arbitrary time-dependent pump $g(t)$, but we first consider solutions for a constant-pump excitation, where $g(t) = g_0 \neq 1$ for $t>0$. The exact solutions are
\begin{gather}
\Delta X^2(t) = \frac{2n_b + 1}{1 + g_0} + \left( \Delta X^2(0) - \frac{2n_b + 1}{1 + g_0}\right) e^{-\Gamma(1 + g_0)t} \label{constx},\\
\Delta Y^2(t) = \frac{2n_b + 1}{1 - g_0} + \left( \Delta Y^2(0) - \frac{2n_b + 1}{1 - g_0}\right) e^{-\Gamma(1 - g_0)t} \label{consty}. 
\end{gather}
When the system is prepared as a thermal state in equilibrium with the bath, both quadratures start with a value of $2n_b + 1$, and \cref{constx,consty} simplify to
\begin{gather}
\Delta X^2(t) = \frac{2n_b + 1}{1 + g_0}\left[ 1 + g_0 e^{-\Gamma(1 + g_0)t}\right] \label{easX}, \\
\Delta Y^2(t) = \frac{2n_b + 1}{1 - g_0}\left[1 - g_0 e^{-\Gamma(1 - g_0)t}\right] \label{easY}.
\end{gather}

Increased squeezing in $X$ will usually be accompanied by increased anti-squeezing in $Y$. Because of this, it is important to examine the merits of using a short, strong pump pulse rather than a long, weak pulse to squeeze the signal.  To this end, we consider the antisqueezing at the time $\tau_1$ at which the squeezing in $X$ reaches the threshold value of $\Delta X^2 = 1$.  For $n_{th}(0)=n_b>0$, we have from \cref{easX,easY} that
\begin{gather}
\Delta Y^2(\tau_1) = \frac{2n_b + 1}{1 - g_0} \left[1 - \left( \frac{1}{g_0} \frac{g_0 - 2n_b}{2 n_b + 1}\right)^{\frac{1 - g_0}{1 + g_0}}\right].
\end{gather}
In the limit of weak pumping ($g_0 \ll 1$), $\Delta Y^2(\tau_1) \rightarrow 2n_b/g_0$, while in the strong pumping limit ($g_0 >> 1$), $\Delta Y^2(\tau_1) \rightarrow 2n_b(2n_b + 1)/(g_0 - 2n_b)$. In both limits, increasing $g_0$ reduces the uncertainty in $\Delta Y^2$ at $t = \tau_1$. Thus, in order to avoid excess growth in the anti-squeezed uncertainty, it is always beneficial to maximize the pumping strength if a specified squeezing is desired. This is because the longer it takes to reach a desired squeezing level, the more thermal photons will be generated due to loss.

We can see from \cref{xDE} that the squeezed quadrature will reach a steady-state value of
\begin{gather}
\Delta X_{min}^2 = \frac{2n_b + 1}{1 + g_0}, \label{minX}
\end{gather} 
for all values of $g_0$. The anti-squeezing, however, will only reach steady state for $g_0<1$, diverging otherwise. Below critical pumping, this anti-squeezing maximum is 
\begin{gather}
\Delta Y^2_{max} = \frac{2n_b + 1}{1 - g_0}.
\end{gather}
In the steady-state, the squeezing is limited to a minimum quadrature uncertainty of $(2n_b + 1)/2$, and therefore we cannot achieve any squeezing if $n_b \geq 0.5$, which agrees with the results for thermalized squeezed states by previous authors \cite{Fearn1988}.

\begin{figure}[ht]
\includegraphics[width=\columnwidth]{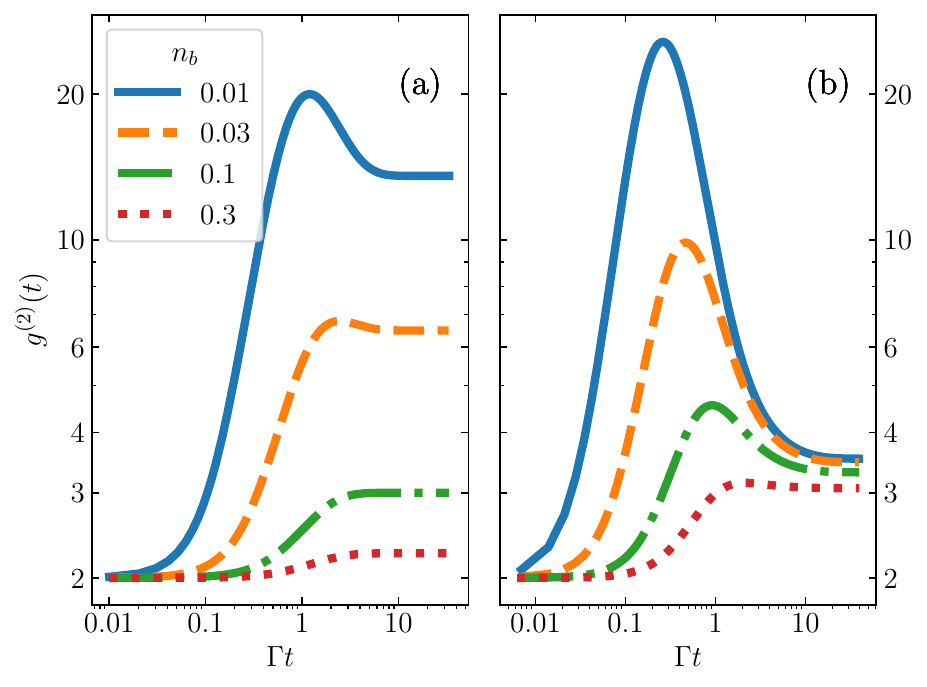}
\caption{Second order simultaneous coherence function $g^{(2)}(t)$ as a function of time for $n_{th}(0)=n_b>0$ for different bath populations, $n_b$. The pump starts at $t=0$ and has a constant amplitude (a) $g_0 = 0.2$ and (b) $g_0 = 0.8$.}
\label{g2fig}
\end{figure}

We now consider the evolution of the equal-time, second order quantum coherence function \cite{WMQO}
\begin{equation}
g^{(2)}\left(t\right)\equiv \frac{Tr\{b^\dag b^\dag b b\rho\left(t\right)\}}{n^2\left(t\right)},    
\end{equation}
which quantifies the correlation between two \textit{simultaneous} photon measurements at time $t$, where $g^{(2)}(t) > 1$ indicates super-Poissonian statistics and photon bunching. In a thermal state $g^{(2)}=2$, while in a STS \cite{properties}
\begin{gather}
g^{(2)} = 2 + \frac{(2n_{th} + 1)^2 \sinh^2(2u)}{[(2n_{th} + 1)\cosh(2u) - 1]^2}. \label{g2}
\end{gather}
In the large population limit ($n \rightarrow \infty$), the STS value approaches the squeezed vacuum state coherence $g^{(2)} = 3$ \cite{hossein}.
In \cref{g2fig}, we plot $g^{(2)}(t)$ as a function of time for two different constant pumping strengths and several different bath temperatures, all with the initial condition $n_{th}(0)=n_b$. We see that $g^{(2)}$ peaks at early times before settling down to a lower steady-state value. As the environmental population is increased, the peak and steady-state values are decreased and the peak occurs at a later time. When the pump is increased, the peak is larger and occurs at an earlier time, while the steady state value is reduced. 
Using \cref{g2,uss,nthss}, we determine the steady-state coherence to be
\begin{gather}
g^{(2)}_{ss} = 2 + \left( \frac{(2n_b + 1)g_0}{2n_b + g_0^2} \right)^2.\label{g2ss}
\end{gather}
The coherence peak above the steady-state value arises as the state transitions from a thermal state to a STS with higher steady-state coherence. As the pumping begins, if $n_b \ll 1$, the population is still small but many squeezed pairs are being created before they can be removed by loss and before the total population becomes too large; because $g^{(2)}$ is normalized to the square of the total population, this results in a larger $g^{(2)}$ during this time period. When the bath temperature is larger though, $g^{(2)}$ is suppressed by the existing thermal population, so we find that the peak is significantly reduced or even absent. 

\begin{figure}[ht]
\includegraphics[width=\columnwidth]{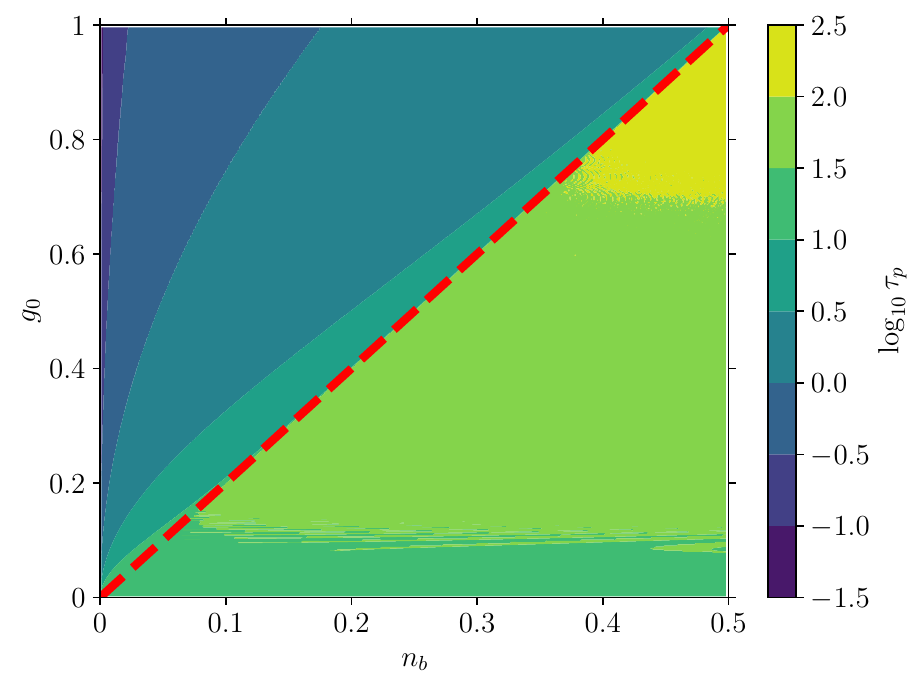}
\caption{Time, $\tau_p$ for the coherence function $g^{2}(t)$ to reach a maximum value as a function of $n_b$ and $g_0$ for a constant pump and initial condition $n_{th}(0)=n_b$. The red dashed line indicates $g_0 = 2n_b$. Note that the peak time is on a logarithmic scale.
}\label{g2peak}
\end{figure}

We can determine when the transition to an STS fails to create a peak in $g^{(2)}(t)$ by finding the time when the coherence is maximized. In \cref{g2peak}, we plot this peak time, $\tau_p$ as a function of the pump strength and bath population. 
We note a clear distinction between the coherence peak times when squeezing is possible ($g_0 > 2n_b$, upper left) and when it is not (lower right). When squeezing is not possible, the coherence reaches the steady-state value monotonically and does not peak. 

\begin{figure}[hbt]
\includegraphics[width=\columnwidth]{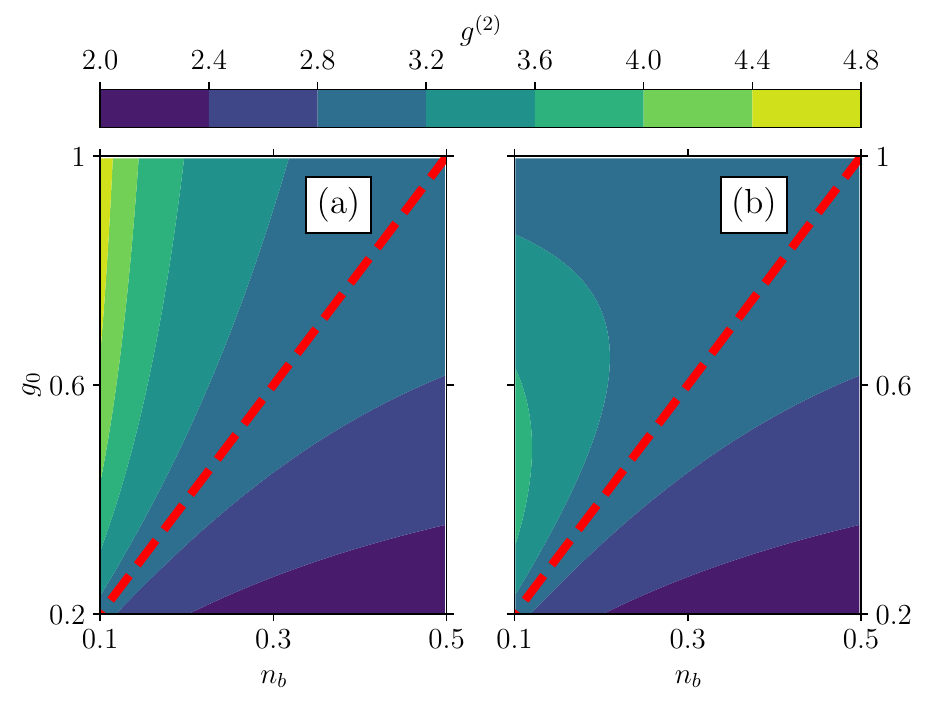}
\caption{Maximum (a) and steady-state (b) second order coherence function, $g^{(2)}$ as a function of $n_b$ and $g_0$ for a constant pump and initial condition $n_{th}(0)=n_b$. Note that the two plots are almost identical below the red dashed line where $g_0 < 2n_b$.}
\label{coherences}
\end{figure}

We have seen from \cref{g2ss} that the steady state coherence will decrease with larger pumping, while the peak value will increase. In \cref{coherences}, we compare the maximum and steady-state coherence as a function of pump strength and bath population. Again, we see that for $g_0 < 2n_b$, the peak value of $g^{(2)}$ is nearly identical to the steady-state value \footnote{Direct comparison between the theoretical steady-state value in \cref{g2ss} and the maximum of the numeric simulation shows a difference of less than $10^{-8}$ for the region $g_0 < 2n_b$, which can be attributed to the limits of the computational precision.}.

\textit{Arbitrary Pump Pulse.}
We now examine the solutions for an arbitrary pump envelope and for a Gaussian pulse. The linear, non-homogeneous ODE in \cref{xDE} has closed-form solution \cite{diffEqns}, 
\begin{equation}
\Delta X^2(t) = \left[ \Gamma(2n_b + 1)\int_0^t q(\tilde{t}) d\tilde{t} + q(0)\Delta X^2(0)\right] q^{-1}(t),    
\end{equation}
where
\begin{equation}
q(t) \equiv \exp\left(\Gamma\int (1 + g(t)) dt\right).    
\end{equation}
If the initial state is a thermal state in equilibrium with the bath, this simplifies to
\begin{gather}
\Delta X^2 (t) = (2n_b + 1)\left[ \Gamma\int_0^t q(t') dt' + q(0)\right] q^{-1}(t), \label{typX}
\end{gather}
which includes the bath temperature only as a prefactor. 
This means that for the usual initial condition of $n_{th}(0)=n_b$, one only needs to solve a single equation to obtain the time evolution of quadrature variance for all bath temperatures. We note that the expression for $\Delta Y^2 (t)$ is identical to the one for $\Delta X^2 (t)$, but with $g(t)\rightarrow -g(t)$.\\
We now consider the particular example of excitation by a Gaussian pulse, with an envelope given by
\begin{gather}
g(t) = g_0 \exp\left[ -\frac{1}{2} \frac{\Gamma^2(t - t_o)^2}{\sigma^2}\right].
\end{gather}

\begin{figure}[hbt]
\includegraphics[width=\columnwidth]{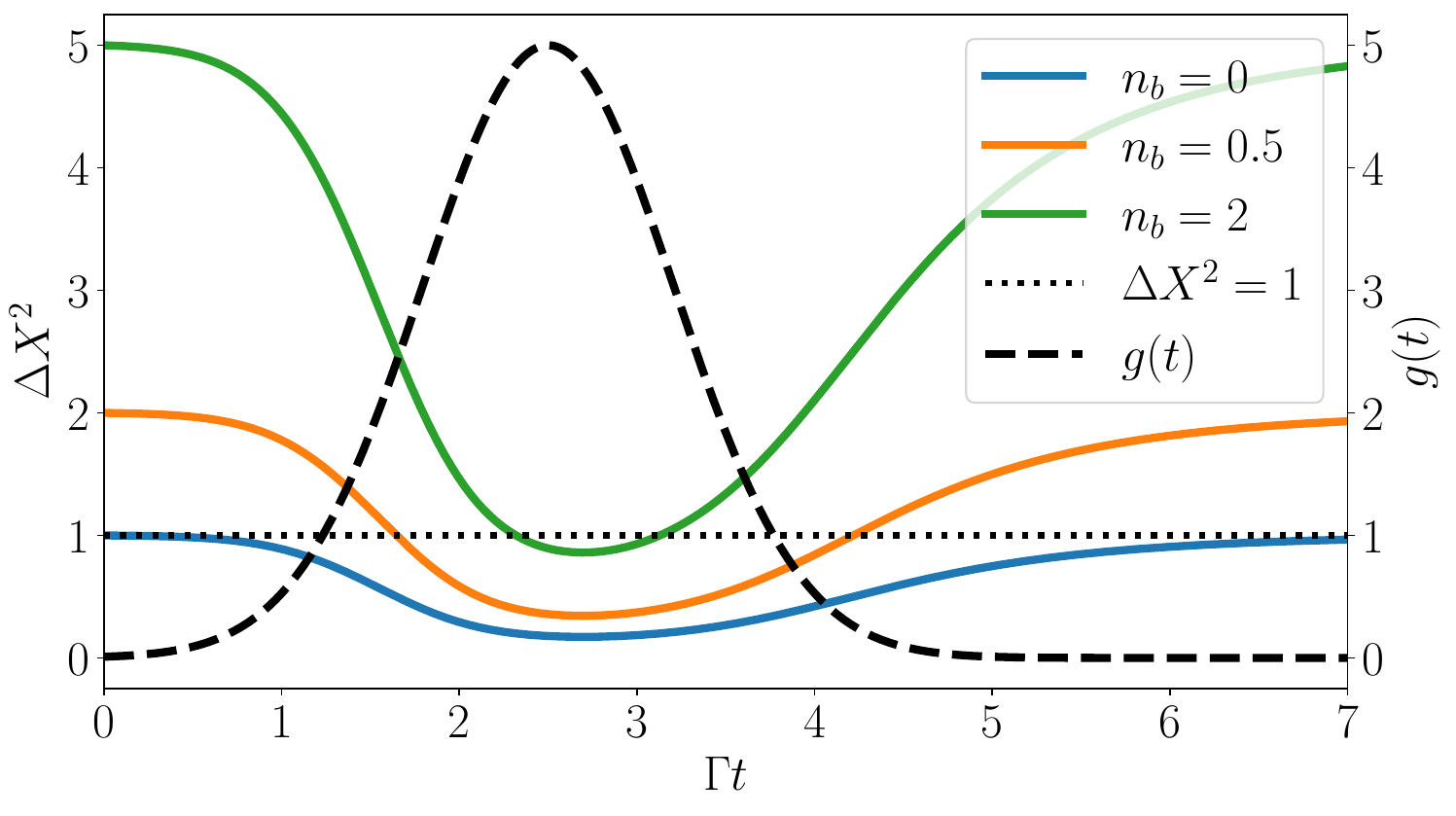}
\caption{Squeezed quadrature $\Delta X^2$ as a function of time for a Gaussian pulse envelope with $g_0$ = 5, $\sigma = 1/\sqrt{2}$, and $\Gamma t_o = 2.5$ for the initial condition $n_{th}(0)=n_b$.}\label{gaussplot}
\end{figure}
In \cref{gaussplot}, we plot the pump envelope as a function of time, along with the squeezed quadrature uncertainty for three different thermal bath populations. We see that the different uncertainty profiles are identical apart from the scaling factor $2n_b + 1$ in \cref{typX}. In particular, the uncertainty is minimized at the same time, $\tau_0$ for all temperatures. Unless the pulse is very short relative to the cavity lifetime, this minimum will occur very close to the time, $\tau_M$ at which the pump reaches its maximum value. Therefore, we can approximate the pump strength at the minimum uncertainty by the maximum pump strength, as was done in Ref. \cite{colinMRR}. Doing this, we obtain
\begin{gather}
\Delta X^2_{min} \approx \frac{2n_b + 1}{1 + g(\tau_M)}.
\end{gather}
Note that for the given pump pulse, the quadrature squeezing below shot noise disappears for $n_b\gtrsim 2.5$. However, with a well-chosen pulse strength and shaping, we see that substantially more squeezing can be achieved than is possible in steady-state. 


\textit{Conclusion.}
In this work, we derived a closed-form solution to the Lindblad master equation, and used it to determine the effects of a thermal environment on the generation of quadrature squeezing via SPDC in a lossy resonator. We proved that the solution is a squeezed thermal state, with contributions to the thermal population coming from loss to and photons from the thermal bath.  We derived a closed-form solution for the evolution of the quadrature uncertainty for an arbitrary un-chirped classical pump pulse and applied it for both a constant and a Gaussian pump pulse.  We found that the thermal bath reduces quadrature squeezing and the equal-time, second order coherence function.


The results presented in this work can be used to help determine the temperature, loss, and pump requirements for squeezed state generation in the few-terahertz regime, where room temperature environmental photons can significantly degrade squeezing. This will be particularly relevant for squeezed state generation in  microwave cavities or optomechanical systems \cite{Han:21,PhysRevLett.95.140504}, where pulse optimization will be necessary to overcome thermal limitations. 
In future work, we plan to extend these results to two-mode cavity system to determine the effect of a thermal environment on the entanglement correlation variance and to directly examine the generation of squeezing in optomechanical systems.

\textit{Acknowledgements}
This work was supported by the Canada Foundation for Innovation and the Natural Sciences and Engineering Research Council of Canada (NSERC).

\bibliography{refs.bib}

\begin{thebibliography}{41}%
\makeatletter
\providecommand \@ifxundefined [1]{%
 \@ifx{#1\undefined}
}%
\providecommand \@ifnum [1]{%
 \ifnum #1\expandafter \@firstoftwo
 \else \expandafter \@secondoftwo
 \fi
}%
\providecommand \@ifx [1]{%
 \ifx #1\expandafter \@firstoftwo
 \else \expandafter \@secondoftwo
 \fi
}%
\providecommand \natexlab [1]{#1}%
\providecommand \enquote  [1]{``#1''}%
\providecommand \bibnamefont  [1]{#1}%
\providecommand \bibfnamefont [1]{#1}%
\providecommand \citenamefont [1]{#1}%
\providecommand \href@noop [0]{\@secondoftwo}%
\providecommand \href [0]{\begingroup \@sanitize@url \@href}%
\providecommand \@href[1]{\@@startlink{#1}\@@href}%
\providecommand \@@href[1]{\endgroup#1\@@endlink}%
\providecommand \@sanitize@url [0]{\catcode `\\12\catcode `\$12\catcode `\&12\catcode `\#12\catcode `\^12\catcode `\_12\catcode `\%12\relax}%
\providecommand \@@startlink[1]{}%
\providecommand \@@endlink[0]{}%
\providecommand \url  [0]{\begingroup\@sanitize@url \@url }%
\providecommand \@url [1]{\endgroup\@href {#1}{\urlprefix }}%
\providecommand \urlprefix  [0]{URL }%
\providecommand \Eprint [0]{\href }%
\providecommand \doibase [0]{https://doi.org/}%
\providecommand \selectlanguage [0]{\@gobble}%
\providecommand \bibinfo  [0]{\@secondoftwo}%
\providecommand \bibfield  [0]{\@secondoftwo}%
\providecommand \translation [1]{[#1]}%
\providecommand \BibitemOpen [0]{}%
\providecommand \bibitemStop [0]{}%
\providecommand \bibitemNoStop [0]{.\EOS\space}%
\providecommand \EOS [0]{\spacefactor3000\relax}%
\providecommand \BibitemShut  [1]{\csname bibitem#1\endcsname}%
\let\auto@bib@innerbib\@empty
\bibitem [{\citenamefont {Milburn}\ and\ \citenamefont {Walls}(1981)}]{milburnOPO}%
  \BibitemOpen
  \bibfield  {author} {\bibinfo {author} {\bibfnamefont {G.}~\bibnamefont {Milburn}}\ and\ \bibinfo {author} {\bibfnamefont {D.}~\bibnamefont {Walls}},\ }\bibfield  {title} {\bibinfo {title} {Production of squeezed states in a degenerate parametric amplifier},\ }\href {https://doi.org/https://doi.org/10.1016/0030-4018(81)90232-7} {\bibfield  {journal} {\bibinfo  {journal} {Optics Communications}\ }\textbf {\bibinfo {volume} {39}},\ \bibinfo {pages} {401} (\bibinfo {year} {1981})}\BibitemShut {NoStop}%
\bibitem [{\citenamefont {Walls}\ and\ \citenamefont {Milburn}(2007)}]{WMQO}%
  \BibitemOpen
  \bibfield  {author} {\bibinfo {author} {\bibfnamefont {D.~F.}\ \bibnamefont {Walls}}\ and\ \bibinfo {author} {\bibfnamefont {G.~J.}\ \bibnamefont {Milburn}},\ }\href {https://doi.org/10.1007/978-3-540-28574-8} {\emph {\bibinfo {title} {{Quantum Optics}}}}\ (\bibinfo  {publisher} {Springer Berlin, Heidelberg},\ \bibinfo {year} {2007})\BibitemShut {NoStop}%
\bibitem [{\citenamefont {Walls}(1983)}]{WallsD.F1983Ssol}%
  \BibitemOpen
  \bibfield  {author} {\bibinfo {author} {\bibfnamefont {D.~F.}\ \bibnamefont {Walls}},\ }\bibfield  {title} {\bibinfo {title} {Squeezed states of light},\ }\href@noop {} {\bibfield  {journal} {\bibinfo  {journal} {Nature (London)}\ }\textbf {\bibinfo {volume} {306}},\ \bibinfo {pages} {141} (\bibinfo {year} {1983})}\BibitemShut {NoStop}%
\bibitem [{\citenamefont {Wu}\ \emph {et~al.}(1986)\citenamefont {Wu}, \citenamefont {Kimble}, \citenamefont {Hall},\ and\ \citenamefont {Wu}}]{PhysRevLett.57.2520}%
  \BibitemOpen
  \bibfield  {author} {\bibinfo {author} {\bibfnamefont {L.-A.}\ \bibnamefont {Wu}}, \bibinfo {author} {\bibfnamefont {H.~J.}\ \bibnamefont {Kimble}}, \bibinfo {author} {\bibfnamefont {J.~L.}\ \bibnamefont {Hall}},\ and\ \bibinfo {author} {\bibfnamefont {H.}~\bibnamefont {Wu}},\ }\bibfield  {title} {\bibinfo {title} {Generation of squeezed states by parametric down conversion},\ }\href {https://doi.org/10.1103/PhysRevLett.57.2520} {\bibfield  {journal} {\bibinfo  {journal} {Phys. Rev. Lett.}\ }\textbf {\bibinfo {volume} {57}},\ \bibinfo {pages} {2520} (\bibinfo {year} {1986})}\BibitemShut {NoStop}%
\bibitem [{\citenamefont {Clerk}\ \emph {et~al.}(2010)\citenamefont {Clerk}, \citenamefont {Devoret}, \citenamefont {Girvin}, \citenamefont {Marquardt},\ and\ \citenamefont {Schoelkopf}}]{RevModPhys.82.1155}%
  \BibitemOpen
  \bibfield  {author} {\bibinfo {author} {\bibfnamefont {A.~A.}\ \bibnamefont {Clerk}}, \bibinfo {author} {\bibfnamefont {M.~H.}\ \bibnamefont {Devoret}}, \bibinfo {author} {\bibfnamefont {S.~M.}\ \bibnamefont {Girvin}}, \bibinfo {author} {\bibfnamefont {F.}~\bibnamefont {Marquardt}},\ and\ \bibinfo {author} {\bibfnamefont {R.~J.}\ \bibnamefont {Schoelkopf}},\ }\bibfield  {title} {\bibinfo {title} {Introduction to quantum noise, measurement, and amplification},\ }\href {https://doi.org/10.1103/RevModPhys.82.1155} {\bibfield  {journal} {\bibinfo  {journal} {Rev. Mod. Phys.}\ }\textbf {\bibinfo {volume} {82}},\ \bibinfo {pages} {1155} (\bibinfo {year} {2010})}\BibitemShut {NoStop}%
\bibitem [{\citenamefont {Kim}\ \emph {et~al.}(2002)\citenamefont {Kim}, \citenamefont {Son}, \citenamefont {Bu\ifmmode~\check{z}\else \v{z}\fi{}ek},\ and\ \citenamefont {Knight}}]{PhysRevA.65.032323}%
  \BibitemOpen
  \bibfield  {author} {\bibinfo {author} {\bibfnamefont {M.~S.}\ \bibnamefont {Kim}}, \bibinfo {author} {\bibfnamefont {W.}~\bibnamefont {Son}}, \bibinfo {author} {\bibfnamefont {V.}~\bibnamefont {Bu\ifmmode~\check{z}\else \v{z}\fi{}ek}},\ and\ \bibinfo {author} {\bibfnamefont {P.~L.}\ \bibnamefont {Knight}},\ }\bibfield  {title} {\bibinfo {title} {Entanglement by a beam splitter: Nonclassicality as a prerequisite for entanglement},\ }\href {https://doi.org/10.1103/PhysRevA.65.032323} {\bibfield  {journal} {\bibinfo  {journal} {Phys. Rev. A}\ }\textbf {\bibinfo {volume} {65}},\ \bibinfo {pages} {032323} (\bibinfo {year} {2002})}\BibitemShut {NoStop}%
\bibitem [{\citenamefont {Azuma}\ \emph {et~al.}(2024)\citenamefont {Azuma}, \citenamefont {Munro},\ and\ \citenamefont {Nemoto}}]{PhysRevA.109.053711}%
  \BibitemOpen
  \bibfield  {author} {\bibinfo {author} {\bibfnamefont {H.}~\bibnamefont {Azuma}}, \bibinfo {author} {\bibfnamefont {W.~J.}\ \bibnamefont {Munro}},\ and\ \bibinfo {author} {\bibfnamefont {K.}~\bibnamefont {Nemoto}},\ }\bibfield  {title} {\bibinfo {title} {Heralded single-photon source based on superpositions of squeezed states},\ }\href {https://doi.org/10.1103/PhysRevA.109.053711} {\bibfield  {journal} {\bibinfo  {journal} {Phys. Rev. A}\ }\textbf {\bibinfo {volume} {109}},\ \bibinfo {pages} {053711} (\bibinfo {year} {2024})}\BibitemShut {NoStop}%
\bibitem [{\citenamefont {Grangier}\ \emph {et~al.}(1987)\citenamefont {Grangier}, \citenamefont {Slusher}, \citenamefont {Yurke},\ and\ \citenamefont {LaPorta}}]{GrangierP.1987Sepi}%
  \BibitemOpen
  \bibfield  {author} {\bibinfo {author} {\bibfnamefont {P.}~\bibnamefont {Grangier}}, \bibinfo {author} {\bibfnamefont {R.~E.}\ \bibnamefont {Slusher}}, \bibinfo {author} {\bibfnamefont {B.}~\bibnamefont {Yurke}},\ and\ \bibinfo {author} {\bibfnamefont {A.}~\bibnamefont {LaPorta}},\ }\bibfield  {title} {\bibinfo {title} {Squeezed-light enhanced polarization interferometer},\ }\href@noop {} {\bibfield  {journal} {\bibinfo  {journal} {Phys. Rev. Lett.}\ }\textbf {\bibinfo {volume} {59}},\ \bibinfo {pages} {2153} (\bibinfo {year} {1987})}\BibitemShut {NoStop}%
\bibitem [{\citenamefont {Aasi}\ \emph {et~al.}(2013)\citenamefont {Aasi} \emph {et~al.}}]{ligo}%
  \BibitemOpen
  \bibfield  {author} {\bibinfo {author} {\bibfnamefont {J.}~\bibnamefont {Aasi}} \emph {et~al.},\ }\bibfield  {title} {\bibinfo {title} {Enhanced sensitivity of the ligo gravitational wave detector by using squeezed states of light},\ }\href {https://doi.org/https://doi.org/10.1038/nphoton.2013.177} {\bibfield  {journal} {\bibinfo  {journal} {Nature Photonics}\ }\textbf {\bibinfo {volume} {7}},\ \bibinfo {pages} {613} (\bibinfo {year} {2013})}\BibitemShut {NoStop}%
\bibitem [{\citenamefont {Furusawa}\ \emph {et~al.}(2015)\citenamefont {Furusawa}, \citenamefont {Masada}, \citenamefont {Kazunori} \emph {et~al.}}]{CVonchip}%
  \BibitemOpen
  \bibfield  {author} {\bibinfo {author} {\bibfnamefont {A.}~\bibnamefont {Furusawa}}, \bibinfo {author} {\bibfnamefont {G.}~\bibnamefont {Masada}}, \bibinfo {author} {\bibnamefont {Kazunori}}, \emph {et~al.},\ }\bibfield  {title} {\bibinfo {title} {Continuous-variable entanglement on a chip},\ }\href {https://doi.org/doi:10.1038/nphoton.2015.42} {\bibfield  {journal} {\bibinfo  {journal} {Nature Photonics}\ }\textbf {\bibinfo {volume} {9}},\ \bibinfo {pages} {316} (\bibinfo {year} {2015})}\BibitemShut {NoStop}%
\bibitem [{\citenamefont {Asavanant}\ and\ \citenamefont {Furusawa}(2024)}]{PhysRevA.109.040101}%
  \BibitemOpen
  \bibfield  {author} {\bibinfo {author} {\bibfnamefont {W.}~\bibnamefont {Asavanant}}\ and\ \bibinfo {author} {\bibfnamefont {A.}~\bibnamefont {Furusawa}},\ }\bibfield  {title} {\bibinfo {title} {Multipartite continuous-variable optical quantum entanglement: Generation and application},\ }\href {https://doi.org/10.1103/PhysRevA.109.040101} {\bibfield  {journal} {\bibinfo  {journal} {Phys. Rev. A}\ }\textbf {\bibinfo {volume} {109}},\ \bibinfo {pages} {040101} (\bibinfo {year} {2024})}\BibitemShut {NoStop}%
\bibitem [{\citenamefont {Vendromin}\ and\ \citenamefont {Dignam}(2020)}]{colinMRR}%
  \BibitemOpen
  \bibfield  {author} {\bibinfo {author} {\bibfnamefont {C.}~\bibnamefont {Vendromin}}\ and\ \bibinfo {author} {\bibfnamefont {M.~M.}\ \bibnamefont {Dignam}},\ }\bibfield  {title} {\bibinfo {title} {Optimization of a lossy microring resonator system for the generation of quadrature-squeezed states},\ }\href {https://doi.org/10.1103/PhysRevA.102.023705} {\bibfield  {journal} {\bibinfo  {journal} {Phys. Rev. A}\ }\textbf {\bibinfo {volume} {102}},\ \bibinfo {pages} {023705} (\bibinfo {year} {2020})}\BibitemShut {NoStop}%
\bibitem [{\citenamefont {Yang}\ \emph {et~al.}(2007)\citenamefont {Yang}, \citenamefont {Chak}, \citenamefont {Bristow}, \citenamefont {van Driel}, \citenamefont {Iyer}, \citenamefont {Aitchison}, \citenamefont {Smirl},\ and\ \citenamefont {Sipe}}]{Yang:07}%
  \BibitemOpen
  \bibfield  {author} {\bibinfo {author} {\bibfnamefont {Z.}~\bibnamefont {Yang}}, \bibinfo {author} {\bibfnamefont {P.}~\bibnamefont {Chak}}, \bibinfo {author} {\bibfnamefont {A.~D.}\ \bibnamefont {Bristow}}, \bibinfo {author} {\bibfnamefont {H.~M.}\ \bibnamefont {van Driel}}, \bibinfo {author} {\bibfnamefont {R.}~\bibnamefont {Iyer}}, \bibinfo {author} {\bibfnamefont {J.~S.}\ \bibnamefont {Aitchison}}, \bibinfo {author} {\bibfnamefont {A.~L.}\ \bibnamefont {Smirl}},\ and\ \bibinfo {author} {\bibfnamefont {J.~E.}\ \bibnamefont {Sipe}},\ }\bibfield  {title} {\bibinfo {title} {Enhanced second-harmonic generation in algaas microring resonators},\ }\href {https://doi.org/10.1364/OL.32.000826} {\bibfield  {journal} {\bibinfo  {journal} {Opt. Lett.}\ }\textbf {\bibinfo {volume} {32}},\ \bibinfo {pages} {826} (\bibinfo {year} {2007})}\BibitemShut {NoStop}%
\bibitem [{\citenamefont {Dutt}\ \emph {et~al.}(2016)\citenamefont {Dutt}, \citenamefont {Miller}, \citenamefont {Luke}, \citenamefont {Cardenas}, \citenamefont {Gaeta}, \citenamefont {Nussenzveig},\ and\ \citenamefont {Lipson}}]{Dutt:16}%
  \BibitemOpen
  \bibfield  {author} {\bibinfo {author} {\bibfnamefont {A.}~\bibnamefont {Dutt}}, \bibinfo {author} {\bibfnamefont {S.}~\bibnamefont {Miller}}, \bibinfo {author} {\bibfnamefont {K.}~\bibnamefont {Luke}}, \bibinfo {author} {\bibfnamefont {J.}~\bibnamefont {Cardenas}}, \bibinfo {author} {\bibfnamefont {A.~L.}\ \bibnamefont {Gaeta}}, \bibinfo {author} {\bibfnamefont {P.}~\bibnamefont {Nussenzveig}},\ and\ \bibinfo {author} {\bibfnamefont {M.}~\bibnamefont {Lipson}},\ }\bibfield  {title} {\bibinfo {title} {Tunable squeezing using coupled ring resonators on a silicon nitride chip},\ }\href {https://doi.org/10.1364/OL.41.000223} {\bibfield  {journal} {\bibinfo  {journal} {Opt. Lett.}\ }\textbf {\bibinfo {volume} {41}},\ \bibinfo {pages} {223} (\bibinfo {year} {2016})}\BibitemShut {NoStop}%
\bibitem [{\citenamefont {Yurke}(1984)}]{Yurke1984}%
  \BibitemOpen
  \bibfield  {author} {\bibinfo {author} {\bibfnamefont {B.}~\bibnamefont {Yurke}},\ }\bibfield  {title} {\bibinfo {title} {Use of cavities in squeezed-state generation},\ }\href@noop {} {\bibfield  {journal} {\bibinfo  {journal} {Phys. Rev. A}\ }\textbf {\bibinfo {volume} {29}} (\bibinfo {year} {1984})}\BibitemShut {NoStop}%
\bibitem [{\citenamefont {Seifoory}\ \emph {et~al.}(2017)\citenamefont {Seifoory}, \citenamefont {Doutre}, \citenamefont {Dignam},\ and\ \citenamefont {Sipe}}]{hossein}%
  \BibitemOpen
  \bibfield  {author} {\bibinfo {author} {\bibfnamefont {H.}~\bibnamefont {Seifoory}}, \bibinfo {author} {\bibfnamefont {S.}~\bibnamefont {Doutre}}, \bibinfo {author} {\bibfnamefont {M.~M.}\ \bibnamefont {Dignam}},\ and\ \bibinfo {author} {\bibfnamefont {J.~E.}\ \bibnamefont {Sipe}},\ }\bibfield  {title} {\bibinfo {title} {Squeezed thermal states: the result of parametric down conversion in lossy cavities},\ }\href {https://doi.org/https://doi.org/10.1364/JOSAB.34.001587} {\bibfield  {journal} {\bibinfo  {journal} {Journal of the Optical Society of America B}\ }\textbf {\bibinfo {volume} {34}},\ \bibinfo {pages} {1587} (\bibinfo {year} {2017})}\BibitemShut {NoStop}%
\bibitem [{\citenamefont {Schneider}\ \emph {et~al.}(1988)\citenamefont {Schneider}, \citenamefont {Lang}, \citenamefont {Mlynek}, \citenamefont {Schiller}, \citenamefont {Marin}, \citenamefont {Bramati}, \citenamefont {Giacobino}, \citenamefont {c~Zhang}, \citenamefont {Poizat}, \citenamefont {f~Roch},\ and\ \citenamefont {Grangier}}]{Schneider1987}%
  \BibitemOpen
  \bibfield  {author} {\bibinfo {author} {\bibfnamefont {K.}~\bibnamefont {Schneider}}, \bibinfo {author} {\bibfnamefont {M.}~\bibnamefont {Lang}}, \bibinfo {author} {\bibfnamefont {J.}~\bibnamefont {Mlynek}}, \bibinfo {author} {\bibfnamefont {S.}~\bibnamefont {Schiller}}, \bibinfo {author} {\bibfnamefont {F.}~\bibnamefont {Marin}}, \bibinfo {author} {\bibfnamefont {A.}~\bibnamefont {Bramati}}, \bibinfo {author} {\bibfnamefont {E.}~\bibnamefont {Giacobino}}, \bibinfo {author} {\bibfnamefont {T.}~\bibnamefont {c~Zhang}}, \bibinfo {author} {\bibfnamefont {J.-P.}\ \bibnamefont {Poizat}}, \bibinfo {author} {\bibfnamefont {J.}~\bibnamefont {f~Roch}},\ and\ \bibinfo {author} {\bibfnamefont {P.}~\bibnamefont {Grangier}},\ }\bibfield  {title} {\bibinfo {title} {Generation of strongly squeezed continuous-wave light at 1064 nm},\ }\href {http://quantum-optics.physik.uni-konstanz.de} {\bibfield  {journal} {\bibinfo  {journal} {Opt. Express}\ }\textbf {\bibinfo {volume} {2}},\ \bibinfo {pages} {59} (\bibinfo {year}
  {1988})}\BibitemShut {NoStop}%
\bibitem [{\citenamefont {Wu}\ \emph {et~al.}(1987)\citenamefont {Wu}, \citenamefont {Xiao},\ and\ \citenamefont {Kimble}}]{Wu1987}%
  \BibitemOpen
  \bibfield  {author} {\bibinfo {author} {\bibfnamefont {L.-A.}\ \bibnamefont {Wu}}, \bibinfo {author} {\bibfnamefont {M.}~\bibnamefont {Xiao}},\ and\ \bibinfo {author} {\bibfnamefont {H.~J.}\ \bibnamefont {Kimble}},\ }\bibfield  {title} {\bibinfo {title} {Squeezed states of light from an optical parametric oscillator},\ }\href@noop {} {\bibfield  {journal} {\bibinfo  {journal} {J. Opt. Soc. Am. B}\ }\textbf {\bibinfo {volume} {4}} (\bibinfo {year} {1987})}\BibitemShut {NoStop}%
\bibitem [{\citenamefont {Aoki}\ \emph {et~al.}(2006)\citenamefont {Aoki}, \citenamefont {Takahashi},\ and\ \citenamefont {Furusawa}}]{Aoki2006}%
  \BibitemOpen
  \bibfield  {author} {\bibinfo {author} {\bibfnamefont {T.}~\bibnamefont {Aoki}}, \bibinfo {author} {\bibfnamefont {G.}~\bibnamefont {Takahashi}},\ and\ \bibinfo {author} {\bibfnamefont {A.}~\bibnamefont {Furusawa}},\ }\bibfield  {title} {\bibinfo {title} {Squeezing at 946nm with periodically poled ktipo(4)},\ }\href@noop {} {\bibfield  {journal} {\bibinfo  {journal} {Optics Epress}\ }\textbf {\bibinfo {volume} {14}},\ \bibinfo {pages} {6930} (\bibinfo {year} {2006})}\BibitemShut {NoStop}%
\bibitem [{\citenamefont {Takeno}\ \emph {et~al.}(2007)\citenamefont {Takeno}, \citenamefont {Yukawa}, \citenamefont {Yonezawa},\ and\ \citenamefont {Furusawa}}]{Takeno2007}%
  \BibitemOpen
  \bibfield  {author} {\bibinfo {author} {\bibfnamefont {Y.}~\bibnamefont {Takeno}}, \bibinfo {author} {\bibfnamefont {M.}~\bibnamefont {Yukawa}}, \bibinfo {author} {\bibfnamefont {H.}~\bibnamefont {Yonezawa}},\ and\ \bibinfo {author} {\bibfnamefont {A.}~\bibnamefont {Furusawa}},\ }\bibfield  {title} {\bibinfo {title} {Observation of -9 db quadrature squeezing with improvement of phase stability in homodyne measurement},\ }\href {https://doi.org/10.1364/OE.15.004321} {\bibfield  {journal} {\bibinfo  {journal} {Opt. Express}\ }\textbf {\bibinfo {volume} {15}},\ \bibinfo {pages} {4321} (\bibinfo {year} {2007})}\BibitemShut {NoStop}%
\bibitem [{\citenamefont {Vahlbruch}\ \emph {et~al.}(2008)\citenamefont {Vahlbruch}, \citenamefont {Mehmet}, \citenamefont {Chelkowski}, \citenamefont {Hage}, \citenamefont {Franzen}, \citenamefont {Lastzka}, \citenamefont {Goßler}, \citenamefont {Danzmann},\ and\ \citenamefont {Schnabel}}]{Vahlbruch2008}%
  \BibitemOpen
  \bibfield  {author} {\bibinfo {author} {\bibfnamefont {H.}~\bibnamefont {Vahlbruch}}, \bibinfo {author} {\bibfnamefont {M.}~\bibnamefont {Mehmet}}, \bibinfo {author} {\bibfnamefont {S.}~\bibnamefont {Chelkowski}}, \bibinfo {author} {\bibfnamefont {B.}~\bibnamefont {Hage}}, \bibinfo {author} {\bibfnamefont {A.}~\bibnamefont {Franzen}}, \bibinfo {author} {\bibfnamefont {N.}~\bibnamefont {Lastzka}}, \bibinfo {author} {\bibfnamefont {S.}~\bibnamefont {Goßler}}, \bibinfo {author} {\bibfnamefont {K.}~\bibnamefont {Danzmann}},\ and\ \bibinfo {author} {\bibfnamefont {R.}~\bibnamefont {Schnabel}},\ }\bibfield  {title} {\bibinfo {title} {Observation of squeezed light with 10-db quantum-noise reduction},\ }\href {https://doi.org/10.1103/PhysRevLett.100.033602} {\bibfield  {journal} {\bibinfo  {journal} {Phys. Rev. Lett.}\ }\textbf {\bibinfo {volume} {100}},\ \bibinfo {pages} {033602} (\bibinfo {year} {2008})}\BibitemShut {NoStop}%
\bibitem [{\citenamefont {Mehmet}\ \emph {et~al.}(2011)\citenamefont {Mehmet}, \citenamefont {Ast}, \citenamefont {Eberle}, \citenamefont {Steinlechner}, \citenamefont {Vahlbruch},\ and\ \citenamefont {Schnabel}}]{Mehmet2011}%
  \BibitemOpen
  \bibfield  {author} {\bibinfo {author} {\bibfnamefont {M.}~\bibnamefont {Mehmet}}, \bibinfo {author} {\bibfnamefont {S.}~\bibnamefont {Ast}}, \bibinfo {author} {\bibfnamefont {T.}~\bibnamefont {Eberle}}, \bibinfo {author} {\bibfnamefont {S.}~\bibnamefont {Steinlechner}}, \bibinfo {author} {\bibfnamefont {H.}~\bibnamefont {Vahlbruch}},\ and\ \bibinfo {author} {\bibfnamefont {R.}~\bibnamefont {Schnabel}},\ }\bibfield  {title} {\bibinfo {title} {Quantum optics; (270.6570) squeezed states; (190.4970) parametric oscillators and amplifiers},\ }\href {https://tds.ego-gw.it/ql/?c=6589.} {\bibfield  {journal} {\bibinfo  {journal} {Optics Epress}\ }\textbf {\bibinfo {volume} {19}},\ \bibinfo {pages} {25763} (\bibinfo {year} {2011})}\BibitemShut {NoStop}%
\bibitem [{\citenamefont {Caves}\ \emph {et~al.}(1980)\citenamefont {Caves}, \citenamefont {Thorne}, \citenamefont {Drever}, \citenamefont {Sandberg},\ and\ \citenamefont {Zimmermann}}]{RevModPhys.52.341}%
  \BibitemOpen
  \bibfield  {author} {\bibinfo {author} {\bibfnamefont {C.~M.}\ \bibnamefont {Caves}}, \bibinfo {author} {\bibfnamefont {K.~S.}\ \bibnamefont {Thorne}}, \bibinfo {author} {\bibfnamefont {R.~W.~P.}\ \bibnamefont {Drever}}, \bibinfo {author} {\bibfnamefont {V.~D.}\ \bibnamefont {Sandberg}},\ and\ \bibinfo {author} {\bibfnamefont {M.}~\bibnamefont {Zimmermann}},\ }\bibfield  {title} {\bibinfo {title} {On the measurement of a weak classical force coupled to a quantum-mechanical oscillator. i. issues of principle},\ }\href {https://doi.org/10.1103/RevModPhys.52.341} {\bibfield  {journal} {\bibinfo  {journal} {Rev. Mod. Phys.}\ }\textbf {\bibinfo {volume} {52}},\ \bibinfo {pages} {341} (\bibinfo {year} {1980})}\BibitemShut {NoStop}%
\bibitem [{\citenamefont {Ruskov}\ \emph {et~al.}(2005)\citenamefont {Ruskov}, \citenamefont {Schwab},\ and\ \citenamefont {Korotkov}}]{PhysRevB.71.235407}%
  \BibitemOpen
  \bibfield  {author} {\bibinfo {author} {\bibfnamefont {R.}~\bibnamefont {Ruskov}}, \bibinfo {author} {\bibfnamefont {K.}~\bibnamefont {Schwab}},\ and\ \bibinfo {author} {\bibfnamefont {A.~N.}\ \bibnamefont {Korotkov}},\ }\bibfield  {title} {\bibinfo {title} {Squeezing of a nanomechanical resonator by quantum nondemolition measurement and feedback},\ }\href {https://doi.org/10.1103/PhysRevB.71.235407} {\bibfield  {journal} {\bibinfo  {journal} {Phys. Rev. B}\ }\textbf {\bibinfo {volume} {71}},\ \bibinfo {pages} {235407} (\bibinfo {year} {2005})}\BibitemShut {NoStop}%
\bibitem [{\citenamefont {Lei}\ \emph {et~al.}(2016)\citenamefont {Lei}, \citenamefont {Weinstein}, \citenamefont {Suh}, \citenamefont {Wollman}, \citenamefont {Kronwald}, \citenamefont {Marquardt}, \citenamefont {Clerk},\ and\ \citenamefont {Schwab}}]{PhysRevLett.117.100801}%
  \BibitemOpen
  \bibfield  {author} {\bibinfo {author} {\bibfnamefont {C.~U.}\ \bibnamefont {Lei}}, \bibinfo {author} {\bibfnamefont {A.~J.}\ \bibnamefont {Weinstein}}, \bibinfo {author} {\bibfnamefont {J.}~\bibnamefont {Suh}}, \bibinfo {author} {\bibfnamefont {E.~E.}\ \bibnamefont {Wollman}}, \bibinfo {author} {\bibfnamefont {A.}~\bibnamefont {Kronwald}}, \bibinfo {author} {\bibfnamefont {F.}~\bibnamefont {Marquardt}}, \bibinfo {author} {\bibfnamefont {A.~A.}\ \bibnamefont {Clerk}},\ and\ \bibinfo {author} {\bibfnamefont {K.~C.}\ \bibnamefont {Schwab}},\ }\bibfield  {title} {\bibinfo {title} {Quantum nondemolition measurement of a quantum squeezed state beyond the 3 db limit},\ }\href {https://doi.org/10.1103/PhysRevLett.117.100801} {\bibfield  {journal} {\bibinfo  {journal} {Phys. Rev. Lett.}\ }\textbf {\bibinfo {volume} {117}},\ \bibinfo {pages} {100801} (\bibinfo {year} {2016})}\BibitemShut {NoStop}%
\bibitem [{\citenamefont {Kronwald}\ \emph {et~al.}(2013)\citenamefont {Kronwald}, \citenamefont {Marquardt},\ and\ \citenamefont {Clerk}}]{KronwaldAndreas2013Alsb}%
  \BibitemOpen
  \bibfield  {author} {\bibinfo {author} {\bibfnamefont {A.}~\bibnamefont {Kronwald}}, \bibinfo {author} {\bibfnamefont {F.}~\bibnamefont {Marquardt}},\ and\ \bibinfo {author} {\bibfnamefont {A.~A.}\ \bibnamefont {Clerk}},\ }\bibfield  {title} {\bibinfo {title} {Arbitrarily large steady-state bosonic squeezing via dissipation},\ }\href@noop {} {\bibfield  {journal} {\bibinfo  {journal} {Phys. Rev. A}\ }\textbf {\bibinfo {volume} {88}},\ \bibinfo {pages} {063833} (\bibinfo {year} {2013})}\BibitemShut {NoStop}%
\bibitem [{\citenamefont {Dassonneville}\ \emph {et~al.}(2021)\citenamefont {Dassonneville}, \citenamefont {Assouly}, \citenamefont {Peronnin}, \citenamefont {Clerk}, \citenamefont {Bienfait},\ and\ \citenamefont {Huard}}]{PRXQuantum.2.020323}%
  \BibitemOpen
  \bibfield  {author} {\bibinfo {author} {\bibfnamefont {R.}~\bibnamefont {Dassonneville}}, \bibinfo {author} {\bibfnamefont {R.}~\bibnamefont {Assouly}}, \bibinfo {author} {\bibfnamefont {T.}~\bibnamefont {Peronnin}}, \bibinfo {author} {\bibfnamefont {A.}~\bibnamefont {Clerk}}, \bibinfo {author} {\bibfnamefont {A.}~\bibnamefont {Bienfait}},\ and\ \bibinfo {author} {\bibfnamefont {B.}~\bibnamefont {Huard}},\ }\bibfield  {title} {\bibinfo {title} {Dissipative stabilization of squeezing beyond 3 db in a microwave mode},\ }\href {https://doi.org/10.1103/PRXQuantum.2.020323} {\bibfield  {journal} {\bibinfo  {journal} {PRX Quantum}\ }\textbf {\bibinfo {volume} {2}},\ \bibinfo {pages} {020323} (\bibinfo {year} {2021})}\BibitemShut {NoStop}%
\bibitem [{\citenamefont {Drummond}\ \emph {et~al.}(1981)\citenamefont {Drummond}, \citenamefont {McNeil},\ and\ \citenamefont {Walls}}]{doi:10.1080/713820531}%
  \BibitemOpen
  \bibfield  {author} {\bibinfo {author} {\bibfnamefont {P.}~\bibnamefont {Drummond}}, \bibinfo {author} {\bibfnamefont {K.}~\bibnamefont {McNeil}},\ and\ \bibinfo {author} {\bibfnamefont {D.}~\bibnamefont {Walls}},\ }\bibfield  {title} {\bibinfo {title} {Non-equilibrium transitions in sub/second harmonic generation},\ }\href {https://doi.org/10.1080/713820531} {\bibfield  {journal} {\bibinfo  {journal} {Optica Acta: International Journal of Optics}\ }\textbf {\bibinfo {volume} {28}},\ \bibinfo {pages} {211} (\bibinfo {year} {1981})}\BibitemShut {NoStop}%
\bibitem [{\citenamefont {Anwar}\ and\ \citenamefont {Zubairy}(1992)}]{PhysRevA.45.1804}%
  \BibitemOpen
  \bibfield  {author} {\bibinfo {author} {\bibfnamefont {J.}~\bibnamefont {Anwar}}\ and\ \bibinfo {author} {\bibfnamefont {M.~S.}\ \bibnamefont {Zubairy}},\ }\bibfield  {title} {\bibinfo {title} {Effect of squeezing on the degenerate parametric oscillator},\ }\href {https://doi.org/10.1103/PhysRevA.45.1804} {\bibfield  {journal} {\bibinfo  {journal} {Phys. Rev. A}\ }\textbf {\bibinfo {volume} {45}},\ \bibinfo {pages} {1804} (\bibinfo {year} {1992})}\BibitemShut {NoStop}%
\bibitem [{\citenamefont {Collett}\ and\ \citenamefont {Gardiner}(1984)}]{PhysRevA.30.1386}%
  \BibitemOpen
  \bibfield  {author} {\bibinfo {author} {\bibfnamefont {M.~J.}\ \bibnamefont {Collett}}\ and\ \bibinfo {author} {\bibfnamefont {C.~W.}\ \bibnamefont {Gardiner}},\ }\bibfield  {title} {\bibinfo {title} {Squeezing of intracavity and traveling-wave light fields produced in parametric amplification},\ }\href {https://doi.org/10.1103/PhysRevA.30.1386} {\bibfield  {journal} {\bibinfo  {journal} {Phys. Rev. A}\ }\textbf {\bibinfo {volume} {30}},\ \bibinfo {pages} {1386} (\bibinfo {year} {1984})}\BibitemShut {NoStop}%
\bibitem [{\citenamefont {Dunlop}\ \emph {et~al.}(2006)\citenamefont {Dunlop}, \citenamefont {Huntington}, \citenamefont {Harb},\ and\ \citenamefont {Ralph}}]{DunlopA.E.2006Goaf}%
  \BibitemOpen
  \bibfield  {author} {\bibinfo {author} {\bibfnamefont {A.~E.}\ \bibnamefont {Dunlop}}, \bibinfo {author} {\bibfnamefont {E.~H.}\ \bibnamefont {Huntington}}, \bibinfo {author} {\bibfnamefont {C.~C.}\ \bibnamefont {Harb}},\ and\ \bibinfo {author} {\bibfnamefont {T.~C.}\ \bibnamefont {Ralph}},\ }\bibfield  {title} {\bibinfo {title} {Generation of a frequency comb of squeezing in an optical parametric oscillator},\ }\href@noop {} {\bibfield  {journal} {\bibinfo  {journal} {Phys. Rev. A}\ }\textbf {\bibinfo {volume} {73}},\ \bibinfo {pages} {013817} (\bibinfo {year} {2006})}\BibitemShut {NoStop}%
\bibitem [{\citenamefont {Jabri}\ and\ \citenamefont {Eleuch}(2019)}]{Jabri:19}%
  \BibitemOpen
  \bibfield  {author} {\bibinfo {author} {\bibfnamefont {H.}~\bibnamefont {Jabri}}\ and\ \bibinfo {author} {\bibfnamefont {H.}~\bibnamefont {Eleuch}},\ }\bibfield  {title} {\bibinfo {title} {Perfect squeezing of terahertz light by two quantum wells using a squeezed vacuum reservoir},\ }\href {https://doi.org/10.1364/JOSAB.36.0000C1} {\bibfield  {journal} {\bibinfo  {journal} {J. Opt. Soc. Am. B}\ }\textbf {\bibinfo {volume} {36}},\ \bibinfo {pages} {C1} (\bibinfo {year} {2019})}\BibitemShut {NoStop}%
\bibitem [{\citenamefont {Breuer}\ and\ \citenamefont {Petruccione}(2007)}]{OpenQuantum}%
  \BibitemOpen
  \bibfield  {author} {\bibinfo {author} {\bibfnamefont {H.-P.}\ \bibnamefont {Breuer}}\ and\ \bibinfo {author} {\bibfnamefont {F.}~\bibnamefont {Petruccione}},\ }\href {https://doi.org/10.1093/acprof:oso/9780199213900.001.0001} {\emph {\bibinfo {title} {{The Theory of Open Quantum Systems}}}}\ (\bibinfo  {publisher} {Oxford University Press},\ \bibinfo {year} {2007})\BibitemShut {NoStop}%
\bibitem [{\citenamefont {Carmicheal}(1999)}]{StatQuantumOptics}%
  \BibitemOpen
  \bibfield  {author} {\bibinfo {author} {\bibfnamefont {H.~J.}\ \bibnamefont {Carmicheal}},\ }\href {https://doi.org/10.1007/978-3-662-03875-8} {\emph {\bibinfo {title} {{Statistical Methods in Quantum Optics}}}}\ (\bibinfo  {publisher} {Springer Berlin, Heidelberg},\ \bibinfo {year} {1999})\BibitemShut {NoStop}%
\bibitem [{Note1()}]{Note1}%
  \BibitemOpen
  \bibinfo {note} {We can see from \protect \cref {nth0ic} that these properties arise not just for an initial state in equilibrium with the thermal bath, but any initial thermal population that satisfies $n_{th}(0) = n_b + a(2n_b + 1)$. The dynamics of the squeezing amplitude will then be independent of the bath, but the decay will scale by the arbitrary factor $a$.}\BibitemShut {Stop}%
\bibitem [{\citenamefont {Kim}\ \emph {et~al.}(1989)\citenamefont {Kim}, \citenamefont {de~Oliveira},\ and\ \citenamefont {Knight}}]{properties}%
  \BibitemOpen
  \bibfield  {author} {\bibinfo {author} {\bibfnamefont {M.~S.}\ \bibnamefont {Kim}}, \bibinfo {author} {\bibfnamefont {F.~A.~M.}\ \bibnamefont {de~Oliveira}},\ and\ \bibinfo {author} {\bibfnamefont {P.~L.}\ \bibnamefont {Knight}},\ }\bibfield  {title} {\bibinfo {title} {Properties of squeezed number states and squeezed thermal states},\ }\href {https://doi.org/10.1103/PhysRevA.40.2494} {\bibfield  {journal} {\bibinfo  {journal} {Phys. Rev. A}\ }\textbf {\bibinfo {volume} {40}},\ \bibinfo {pages} {2494} (\bibinfo {year} {1989})}\BibitemShut {NoStop}%
\bibitem [{\citenamefont {Fearn}\ and\ \citenamefont {Collett}(1988)}]{Fearn1988}%
  \BibitemOpen
  \bibfield  {author} {\bibinfo {author} {\bibfnamefont {H.}~\bibnamefont {Fearn}}\ and\ \bibinfo {author} {\bibfnamefont {M.~J.}\ \bibnamefont {Collett}},\ }\bibfield  {title} {\bibinfo {title} {Representations of squeezed states with thermal noise},\ }\href {https://doi.org/10.1080/09500348814550571} {\bibfield  {journal} {\bibinfo  {journal} {Journal of Modern Optics}\ }\textbf {\bibinfo {volume} {35}},\ \bibinfo {pages} {553} (\bibinfo {year} {1988})}\BibitemShut {NoStop}%
\bibitem [{Note2()}]{Note2}%
  \BibitemOpen
  \bibinfo {note} {Direct comparison between the theoretical steady-state value in \protect \cref {g2ss} and the maximum of the numeric simulation shows a difference of less than $10^{-8}$ for the region $g_0 < 2n_b$, which can be attributed to the limits of the computational precision.}\BibitemShut {Stop}%
\bibitem [{\citenamefont {Constanda}(2017)}]{diffEqns}%
  \BibitemOpen
  \bibfield  {author} {\bibinfo {author} {\bibfnamefont {C.}~\bibnamefont {Constanda}},\ }\href@noop {} {\emph {\bibinfo {title} {Differential equations a primer for scientists and engineers}}},\ \bibinfo {edition} {2nd}\ ed.\ (\bibinfo  {publisher} {Springer Cham},\ \bibinfo {year} {2017})\BibitemShut {NoStop}%
\bibitem [{\citenamefont {Han}\ \emph {et~al.}(2021)\citenamefont {Han}, \citenamefont {Wang},\ and\ \citenamefont {Zhang}}]{Han:21}%
  \BibitemOpen
  \bibfield  {author} {\bibinfo {author} {\bibfnamefont {K.}~\bibnamefont {Han}}, \bibinfo {author} {\bibfnamefont {Y.}~\bibnamefont {Wang}},\ and\ \bibinfo {author} {\bibfnamefont {G.-Q.}\ \bibnamefont {Zhang}},\ }\bibfield  {title} {\bibinfo {title} {Enhancement of microwave squeezing via parametric down-conversion in a superconducting quantum circuit},\ }\href {https://doi.org/10.1364/OE.423373} {\bibfield  {journal} {\bibinfo  {journal} {Opt. Express}\ }\textbf {\bibinfo {volume} {29}},\ \bibinfo {pages} {13451} (\bibinfo {year} {2021})}\BibitemShut {NoStop}%
\bibitem [{\citenamefont {Moon}\ and\ \citenamefont {Girvin}(2005)}]{PhysRevLett.95.140504}%
  \BibitemOpen
  \bibfield  {author} {\bibinfo {author} {\bibfnamefont {K.}~\bibnamefont {Moon}}\ and\ \bibinfo {author} {\bibfnamefont {S.~M.}\ \bibnamefont {Girvin}},\ }\bibfield  {title} {\bibinfo {title} {Theory of microwave parametric down-conversion and squeezing using circuit qed},\ }\href {https://doi.org/10.1103/PhysRevLett.95.140504} {\bibfield  {journal} {\bibinfo  {journal} {Phys. Rev. Lett.}\ }\textbf {\bibinfo {volume} {95}},\ \bibinfo {pages} {140504} (\bibinfo {year} {2005})}\BibitemShut {NoStop}%
\end{thebibliography}%

\setcounter{equation}{0}
\renewcommand{\theequation}{S\arabic{equation}}
\section*{Supplementary Material: Solving the squeezed thermal state equations}
In this supplementary material, we use the technique of Hossein \textit{et. al.} \cite{hossein} to solve equation \cref{Oeqn} with the LME \cref{master}, including the effect of the environmental temperature. We begin by taking the derivative of \cref{Oeqn} and use the chain rule to split it in five parts:
\begin{gather}
\dot{O}(t) = \dot{O}_T(t) + \dot{O}_S(t) + \dot{O}_0 + \dot{O}_V + \dot{O}_L.
\label{odot}
\end{gather}
The first four terms, containing contributions from the thermal state density operator, the squeeze operator, the unperturbed Hamiltonian, and the pump laser are identical to zero-temperature case, derived in the Appendix of Ref. \cite{hossein}. Defining $x \equiv e^{-\hbar\omega/kT}, s \equiv \sinh(u),$ and $c \equiv \cosh(u)$, these expressions for the first four terms can be shown to be
\begin{gather}
\dot{O}_T = \acomm{J}{O}, \label{OT} 
\end{gather}
where
\begin{equation}
J \equiv \frac{d\rho_T^{-1/2}}{dt}\rho_T^{1/2} = \frac{1}{2x}\frac{dx}{dt}(n_{th} - b^\dag b); \label{J}
\end{equation}
\begin{equation}
\dot{O}_S = \acomm{M}{O} + i\comm{N}{O}, \label{OS}
\end{equation}
where
\begin{align}
\begin{aligned}
M + iN = &(-is^2 \dot{\phi})(b^\dag b + \frac{1}{2}) \\+ &\frac{1}{2}\dot{u}(x^{-1}{b^\dag}^2e^{i\phi} - xb^2 e^{-i\phi}) \label{MN}\\
+ &\frac{1}{2}ics\dot{\phi}(x^{-1}{b^\dag}^2e^{i\phi} + xb^2e^{-i\phi});
\end{aligned}
\end{align}
\begin{gather}
\frac{\dot{O}_0}{-i\omega} = GO - OG^\dag = \comm{P}{O} + i\acomm{Q}{O},\label{OO}
\end{gather}
where
\begin{align}
\begin{aligned}
G &\equiv \rho_T^{-1/2}S^\dag b^\dag b S \rho_T^{1/2} \\ &= s^2 + (c^2 + s^2)b^\dag b - cs(x^{-1}{b^\dag}^2e^{i\phi} + x b^2e^{-i\phi}),
\end{aligned}
\end{align}
thus
\begin{align}
\begin{aligned}
P \equiv &s^2 + (c^2 + s^2)b^\dag b \\ - &\frac{1}{2}cs(x^{-1} + x)({b^\dag}^2 e^{i\phi} + b^2 e^{-i\phi}), 
\end{aligned}
\end{align}
\begin{align}
\begin{aligned}
Q \equiv \frac{1}{2}ics(x^{-1} - x)({b^\dag}^2 e^{i\phi} - b^2 e^{-i\phi}); \label{Q}
\end{aligned}
\end{align}
and
\begin{gather}
\dot{O}_V = -\frac{i}{\hbar}\comm{\bar{P}}{O} + \frac{1}{\hbar}\acomm{\bar{Q}}{O}, \label{OV}
\end{gather}
where
\begin{align}
\begin{aligned}
\bar{P} \equiv &-cs(\alpha\gamma e^{-i\phi} + \alpha^*\gamma^*e^{i\phi}) \\ &- 2cs(\alpha\gamma e^{-i\phi} + \alpha^*\gamma^*e^{i\phi}) b^\dag b \\ &+ \frac{1}{2}(\alpha\gamma(x^{-1} + x)c^2 + \alpha^*\gamma^*(x^{-1} + x)s^2e^{2i\phi}){b^\dag}^2\\ &+ \frac{1}{2}(\alpha\gamma(x^{-1} + x)s^2e^{-i\phi} + \alpha^*\gamma^*(x^{-1} + x)c^2)b^2 
\end{aligned}
\end{align}
and
\begin{align}
\begin{aligned}
\bar{Q} = &-\frac{i}{2}(x^{-1} - x)(\alpha^*\gamma^*s^2 e^{2i\phi} + \alpha\gamma c^2){b^\dag}^2 \\ &+ \frac{i}{2}(x^{-1} - x)(\alpha\gamma s^2 e^{2i\phi} + \alpha^*\gamma^* c^2){b^2}. \label{Qbar}
\end{aligned}
\end{align}
The final term $\dot{O}_L$ has new components related to the bath population. We see from the master equation that
\begin{align}
\begin{aligned}
\dot{O}_{L} = \Gamma(n_b + 1) \Big( &\rho_T^{-1/2}S^\dag b\rho b^\dag S \rho_T^{-1/2} \\ &-\frac{1}{2}\rho_T^{-1/2}S^\dag (\acomm{b^\dag b}{\rho}) S \rho_T^{-1/2}\Big) \\
+ \Gamma n_b \Big( &\rho_T^{-1/2}S^\dag b^\dag \rho bS \rho_T^{-1/2} \\ &-\frac{1}{2}\rho_T^{-1/2}S^\dag (\acomm{b b^\dag}{\rho}) S \rho_T^{-1/2}\Big).
\end{aligned} \label{OL}
\end{align}
Now, defining the operators
\begin{gather}
T = \rho_T^{-1/2}S^\dag b S \rho_T^{1/2}, \\
\tilde{T} = \rho_T^{-1/2}S^\dag b^\dag S \rho_T^{1/2},
\end{gather}
and solving
\begin{align}
\begin{aligned}
\tilde{G} & \equiv \rho_T^{-1/2} S^\dag b b^\dag S \rho^{1/2} \\ &= G + \rho_T^{-1/2} S^\dag \comm{b}{b^\dag} S \rho_T^{1/2} = G + 1,
\end{aligned}
\end{align}
\cref{OL} becomes
\begin{align}
\begin{aligned}
\dot{O}_{L} = &\Gamma(n_b + 1) \left[T O T^\dag - \frac{1}{2}(GO + OG^\dag)\right] \\ + &\Gamma n_b \left[\tilde{T} O {\tilde{T}}^\dag -\frac{1}{2}(\tilde{G} O + O{\tilde{G}}^\dag)\right] \\
= & \Gamma(n_b + 1) \left[T O T^\dag - \frac{1}{2}(GO + OG^\dag)\right] \\ + &\Gamma n_b \left[\tilde{T} O {\tilde{T}}^\dag -\frac{1}{2}(GO + OG^\dag) - O\right] \\
=& \Gamma(n_b + 1) \left[T O T^\dag - \frac{1}{2}\acomm{P}{O} - \frac{i}{2}\comm{Q}{O}\right] \\ &+ \Gamma n_b \left[\tilde{T} O {\tilde{T}}^\dag -\frac{1}{2}\acomm{P}{O} - \frac{i}{2}\comm{Q}{O} - O\right].
\end{aligned} \label{OL2}
\end{align}

Inserting \cref{OT,OS,OO,OV,OL2} into \cref{odot} and setting $O$ to the identity for all time yields
\begin{align}
\begin{aligned}
0 &= 2J + 2M + 2\omega Q + \frac{2}{\hbar}\bar{Q} \\ & + \Gamma (n_b + 1)(T T^\dag - P) + \Gamma n_b (\tilde{T} {\tilde{T}}^\dag - P - 1). \label{Oset}
\end{aligned}
\end{align}
solving for the remaining terms,
\begin{gather*}
T T^\dag = xc^2 + (xc^2 + x^{-1}s^2)b^\dag b - cs(b^2 e^{-i\phi} + {b^\dag}^2e^{i\phi}) \\
\tilde{T} {\tilde{T}}^\dag = x^{-1}c^2b^\dag b - cs e^{i\phi}{b^\dag}^2 - cs e^{-i\phi}b^2 + xs^2 (b^\dag b + 1)
\end{gather*}  
we use \cref{J,MN,Q,Qbar} in \cref{Oset}:

\begin{align}
\begin{aligned}
0 &= \frac{1}{x}\frac{dx}{dt}(n_{th} - b^\dag b) 
+ \frac{1}{2} \dot{u}(x^{-1} - x)({b^\dag}^2 e^{i\phi} + b^2 e^{-i\phi}) \\
&+ \frac{1}{2}ics\dot{\phi}(x^{-1} - x)({b^\dag}^2 e^{i\phi} - b^2 e^{-i\phi}) \\
&+ i\omega cs(x^{-1} - x)({b^\dag}^2 e^{i\phi} - b^2 e^{-i\phi}) \\
&+ \frac{i}{\hbar} (x^{-1} - x)(\gamma \alpha s^2 e^{-2i\phi} + \gamma^* \alpha^* c^2)b^2 \\
&- \frac{i}{\hbar}(x^{-1} - x)(\gamma^* \alpha^* s^2 e^{2i\phi} + \gamma \alpha c^2){b^\dag}^2 \\
&\begin{aligned}+ \Gamma(n_b + 1)  \Big(xc^2 &+ (xc^2 + x^{-1}s^2)b^\dag b \\ &-cs(b^2 e^{-i\phi} + {b^\dag}^2 e^{i\phi})\Big)\end{aligned} \\
&\begin{aligned}- \Gamma(n_b + 1)  \Big(s^2 &+ (c^2 + s^2)b^\dag b \\ & - \frac{1}{2}cs(x^{-1} + x)({b^\dag}^2 e^{i\phi} + b^2 e^{-i\phi})\Big) \end{aligned} \\
&\begin{aligned}+ \Gamma n_b  \Big(xs^2 &+ (x^{-1}c^2 + xs^2)b^\dag b 
\\ &-cs(e^{i\phi}{b^\dag}^2 + e^{-i\phi}b^2)\Big) \end{aligned} \\
&\begin{aligned}- \Gamma n_b  \Big(s^2 &+ (c^2 + s^2)b^\dag b \\ &- \frac{1}{2}cs(x^{-1} + x)({b^\dag}^2 e^{i\phi} + b^2 e^{-i\phi}) + 1\Big). \end{aligned} \label{long1}
\end{aligned}
\end{align}

We now introduce the Hermitian operators 
\begin{gather}
\chi_1 = {b^\dag}^2e^{i\phi} + {b}^2e^{-i\phi} \\
\chi_2 = i({b^\dag}^2e^{i\phi} - {b}^2e^{-i\phi}). 
\end{gather} 

Substituting these into \cref{long1}, we obtain
\begin{align}
\begin{aligned}
0 &= \frac{1}{x}\frac{dx}{dt}(n_{th} - b^\dag b) 
+ \frac{1}{2} \dot{u}(x^{-1} - x)\chi_1 \\
&+ \frac{1}{2}cs\dot{\phi}(x^{-1} - x)\chi_2
+ \omega cs(x^{-1} - x)\chi_2 \\
&+ \frac{i}{\hbar} (x^{-1} - x)(\gamma \alpha s^2 e^{-i\phi} + \gamma^* \alpha^* c^2 e^{i\phi})\frac{1}{2}(\chi_1 + i\chi_2) \\
&- \frac{i}{\hbar}(x^{-1} - x)(\gamma^* \alpha^* s^2 e^{i\phi} + \gamma \alpha c^2 e^{-i\phi})\frac{1}{2}(\chi_1 - i\chi_2) \\
&+ \Gamma (n_b + 1) \left(xc^2 + (xc^2 + x^{-1}s^2)b^\dag b - cs\chi_1\right) \\
&- \Gamma (n_b + 1) \left(s^2 + (c^2 + s^2)b^\dag b - \frac{1}{2}cs(x^{-1} + x)\chi_1\right) \\
&+ \Gamma n_b \left(xs^2 + (x^{-1}c^2 + xs^2)b^\dag b - cs\chi_1\right) \\
&- \Gamma n_b \left(s^2 + (c^2 + s^2)b^\dag b - \frac{1}{2}cs(x^{-1} + x)\chi_1\right) \\
&- \Gamma n_b. 
\end{aligned}
\end{align}  

This allows us to form four independent equations using
\begin{gather}
F_1 \chi_1 + F_2 \chi_2 + F_3 b^\dag b + F_4 = 0,
\end{gather}
where
\begin{align}
\begin{aligned}
F_1 &= \frac{1}{2} \dot{u}(x^{-1} - x) + \frac{i}{2\hbar} (x^{-1} - x)(\gamma^* \alpha^* e^{i\phi} - \gamma \alpha e^{-i\phi})\\&+ \Gamma(2n_b + 1)cs\big(\frac{1}{2}(x^{-1} + x) - 1\big),
\end{aligned}
\end{align}
\begin{align}
\begin{aligned}
F_2 &= \frac{1}{2}cs\dot{\phi}(x^{-1} - x) + \omega cs(x^{-1} - x) \\ &- \frac{1}{2\hbar}(x^{-1} - x)(c^2 + s^2)(\gamma^* \alpha^* e^{i\phi} + \gamma \alpha e^{-i\phi}),
\end{aligned}
\end{align}
\begin{align}
\begin{aligned}
F_3 = &-\frac{1}{x}\frac{dx}{dt} + \Gamma (n_b + 1)(xc^2 + x^{-1}s^2 - c^2 - s^2) \\ &+ \Gamma n_b(x^{-1}c^2 + xs^2 - c^2 - s^2),
\end{aligned}
\end{align}
and
\begin{align}
\begin{aligned}
F_4  = & \frac{1}{x}\frac{dx}{dt}n_{th} \\ &+ \Gamma (n_b + 1) (xc^2 - s^2) + \Gamma n_b(xs^2 - s^2 - 1).
\end{aligned}
\end{align}

The equations for $F_3$ and $F_4$ both generate equations in $x$, which are consistent and can be reduced to
\begin{gather}
\frac{dx}{dt} = -(1 - x)\Gamma\left[ n_b(2s^2 + 1)(x-1) + s^2(x-1) + x\right].
\end{gather}
Using $\frac{dn_{th}}{dt} = \frac{1}{(1 - x)^2}\frac{dx}{dt}$, this translates to the following equation for $n_{th}$:
\begin{align}
\begin{aligned}
\frac{dn_{th}}{dt} =& \Gamma\left[n_b(2s^2 + 1) + s^2 - \frac{x}{1-x}\right] \\ =& \Gamma\left[n_b(2s^2 + 1) + s^2 - n_{th} \right],
\end{aligned}
\end{align}
which is equivalent to \cref{nthfirst}. Meanwhile, from $F_2 = 0$, we obtain
\begin{gather}
\dot{\phi} = -2\omega + \frac{1}{\hbar}\frac{c^2 + s^2}{cs}(\gamma^* \alpha^* e^{i\phi} + \gamma \alpha e^{-i\phi}),
\end{gather}
which is \cref{phi0}. Finally, from $F_1 = 0$, we obtain
\begin{gather}
\dot{u} = -\frac{i}{\hbar}(\gamma^* \alpha^* e^{i\phi} - \gamma \alpha e^{-i\phi}) - \frac{\Gamma(2n_b + 1)cs}{2n_{th} + 1},
\end{gather}
which is simply \cref{u0}.

\end{document}